\documentclass[12pt,1p]{elsarticle}
\usepackage{amssymb}
\usepackage{multirow}
\usepackage{amsmath,bm}
\usepackage{mathtools}
\usepackage{textcomp, gensymb}
\usepackage{amsfonts}
\usepackage{caption}
\usepackage{subcaption}
\usepackage{graphicx}
\usepackage{comment}
\usepackage{hyperref}
\usepackage{lineno}
\newcommand{\W}{\Omega}
\newcommand{\ii}{{\rm i}}
\renewcommand{\Re}{{\rm Re}}
\renewcommand{\Im}{{\rm Im}}

\begin{document}
\begin{frontmatter}

\title{Maximal Electromagnetic Coupling \\Between Arbitrary-Shaped Nanotubes}

\author[NU]{Abzal Aznabayev\texorpdfstring{\corref{contrib}}{}}
\author[NU]{Ravil Ashirmametov\texorpdfstring{\corref{contrib}}{}}
\author[NU]{Aidana Faizulla}
\author[NTUA]{Constantinos Valagiannopoulos}
\author[NU]{Konstantinos Kostas}
\affiliation[NU]{organization={Nazarbayev University},
             addressline={School of Engineering and Digital Sciences},
             city={Astana},
             postcode={KZ-010000},
             country={Kazakhstan}}
\affiliation[NTUA]{organization={National Technical University of Athens},
             addressline={School of Electrical and Computer Engineering},
             city={Athens},
             postcode={GR-15780},
             country={Greece}}
%\fntext[equal]{These authors contributed equally to this work.}
\cortext[contrib]{These authors contributed equally to this work.}

\begin{abstract}
The interaction of electromagnetic waves with pairs of nanotubes having ar\-bi\-trary-shaped cross sections is investigated. The study starts by thoroughly examining nanotube couples with circular boundaries to identify the structural, textural and excitation parameters that enhance the electric field concentration. Such an initial step generates a design space across which the regions of interest are determined to formulate a shape optimization problem targeting at the maximization of the signal internally to the nanotubes by modifying their boundary surface while maintaining the values for the remaining operational parameters. An IsoGeometric-Analysis-based Boundary Element Method (IGABEM) is employed for estimating the electric field around nanotube boundaries, which is subsequently used towards the overall solution of the boundary value problem. A combination of global and local optimizers along with geometric parametric models that generate valid nanotube cross sections complement the IGABEM method in the quest of optimal designs. The achieved optimal shapes attain substantial boost in the concentration of the electric field which can easily exceed, by 30 times or more, the corresponding value obtained by circular nanotube pairs. This superior performance is maintained for a large range of wave angles and nanotube areas and, accordingly, the robustness of the obtained results and their potential applicability in a wide range of applications, is demonstrated. Finally, the computational method developed in this work can be easily extended for analyzing a finite array of nanotubes, while insights gained by the reported optimal shapes pave the way for their optimization and fine tuning in collective setups like electromagnetic gratings and photonic metasurfaces.
\end{abstract}

\begin{keyword}
isogeometric analysis\sep boundary element method \sep shape optimization \sep electromagnetic waves \sep carbon nanotubes \sep electric field 

\end{keyword}

\end{frontmatter}
%\begin{linenumbers}
%% Add \usepackage{lineno} before \begin{document} and uncomment 
%% following line to enable line numbers
%% \linenumbers

%% main text
%%
\section{Introduction}\label{sec:intro}
Nanotubes refer to tubular structures with diameters in the nanoscale that are usually formed by thin 2D sheets rolled up to create the corresponding tubular shape. Huge scientific and engineering interest in such nanostructures has been developing since 1991 when Carbon Nanotubes (CNTs) were discovered by Iijima~\cite{Iijima1991}. CNTs are the first generation of nanotubes whose properties sparked the interest in nanotubes exploration due to their amazing mechanical strength, thermal conductivity, and other physical properties that attract attention from scientists and engineers~\cite{POPOV200461}. Because of their high tensile strength and Young's modulus, they are commonly used in composite materials, coatings, and films, to enhance the end-product properties~\cite{doi:10.1126/science.1222453}. Another important property of carbon nanotubes lies in their high electrical conductivity, compared to other materials, which opens a wide range of applications in electronics, electrochemical devices, field emission devices, and nanoscale electric devices~\cite{doi:10.1126/science.1060928}. However, CNTs are not the only type of nanotube studied in the pertinent literature. Boron nitride, gallium nitride, silicon, and carbon-titanium dioxide nanotubes are among the nanotubes which have been considered and studied in pertinent literature~\cite{GOVINDARAJU2014243, PAPANICOLAOU2015735}. 

One of the most intriguing aspects of nanotube properties is their shape dependency, which allows for additional degrees of freedom in their study and deployment. Depending on its circumference curvature, a nanotube might exhibit metallic or semiconducting behavior~\cite{SCHULZ201433}; hence, varying shapes and surface electric conductivities endow significantly different behaviors to nanotubes. For most studies and applications, only circular shapes are considered as their analysis and fabrication are significantly simpler than arbitrary-shaped ones, however, not all nanotubes are perfectly circular. For example, CNTs can be single-walled or multi-walled, with defects and discontinuities present; see~\cite{Charlier2002}. Thus, considering geometric deviations, or more generally arbitrary shapes, offer a new avenue for research in nanotube-properties modifications, which has not been sufficiently studied in pertinent literature. With respect to fabrication of arbitrary-shaped nanotubes, techniques involving long single strands of DNA~\cite{B1-Maune}, standard chemical vapor deposition~\cite{B2-Ren} as well as atomic force microscopy~\cite{B3-AVOURIS1999201} can materialize nanotube surfaces of significant shape complexity.

In this work, we extend the numerical studies performed for the shape optimization of a single nanotube illuminated by electromagnetic (EM) waves~\cite{KOSTAS2023,IZNAT2025106153}, and perform a parametric study for the case of field concentration in a pair of nanotubes with varying separating distances, optical dimensions, and electric surface conductivities, when illuminated by waves with varying incidence angles. Furthermore, for a selected number of cases, we determine the shape of nanotube pairs that lead to electric-field maximization while maintaining the values for the remaining design variables, namely, to more than 30 times enhancement over the circular pair. This study of EM wave interactions with pairs of circular and arbitrary-shaped nanotubes provides us with tools and insights which will pave the way for extending this work to problems dealing with a finite array of nanotubes. The arrays may form the so-called metasurfaces with exotic properties that can be significantly different to the underlying constituent structures~\cite{adma.201003989}. The maximization of the electric field inside nanotubes or nanotube arrays can be proven beneficial for a wide range of applications in communications and data transfer~\cite{Bait-Suwailam19,10.1063/1.4900528,Bonache2016}.

As far as the mathematical formulation is concerned, an IsoGeometric-Analysis-based Boundary Element Method (IGABEM) is employed for the computation of the electric field along the nanotube boundaries which subsequently allows for the signal concentration calculation within the regions enclosed by the nanotubes. The IsoGeometric-Analysis (IGA) method was first introduced by Hughes et al.~\cite{Hughes05, CottrellEtAl09} in the context of finite element analysis and it was later extended to 2D and 3D boundary element approaches (BEM) by various researchers~ \cite{Politis09,Simpson12,Scott2013,GinnisEtAl2014}. Furthermore, the IGABEM approach uses the same basis functions for the representation of the domain geometry and the solution field, which results in several advantages in engineering analysis. In brief, it maintains the accuracy of the geometrical description, eliminates the need for a separate analysis mesh, and significantly reduces the computational effort by requiring a relatively small number of degrees of freedom, compared to standard BEM methods, to achieve accurate representations of the solution~\cite{GinnisEtAl2014,COTTRELL20074160,EVANS20091726}.

The remaining of this paper is structured as follows. In \S\ref{problem formulation}, we formulate the problem for a pair of nanotubes illuminated by planar EM waves and derive the boundary integral equation (BIE) for the calculation of the electric field on the two boundaries. The discretization of the BIE using the IGABEM approach is described in \S\ref{sec:IGABEM}, followed by the description of the parametric study in \S\ref{sec:parametric_study} and the shape optimization problem in~\S\ref{sec:optimization}. The obtained results are discussed in \S\ref{sec:results} with concluding remarks and future research directions summarized in \S\ref{sec:conclusion}.
  
\begin{figure}
    \centering
    \begin{subfigure}[t]{0.49\textwidth}
        \centering
        \includegraphics[width=\textwidth]{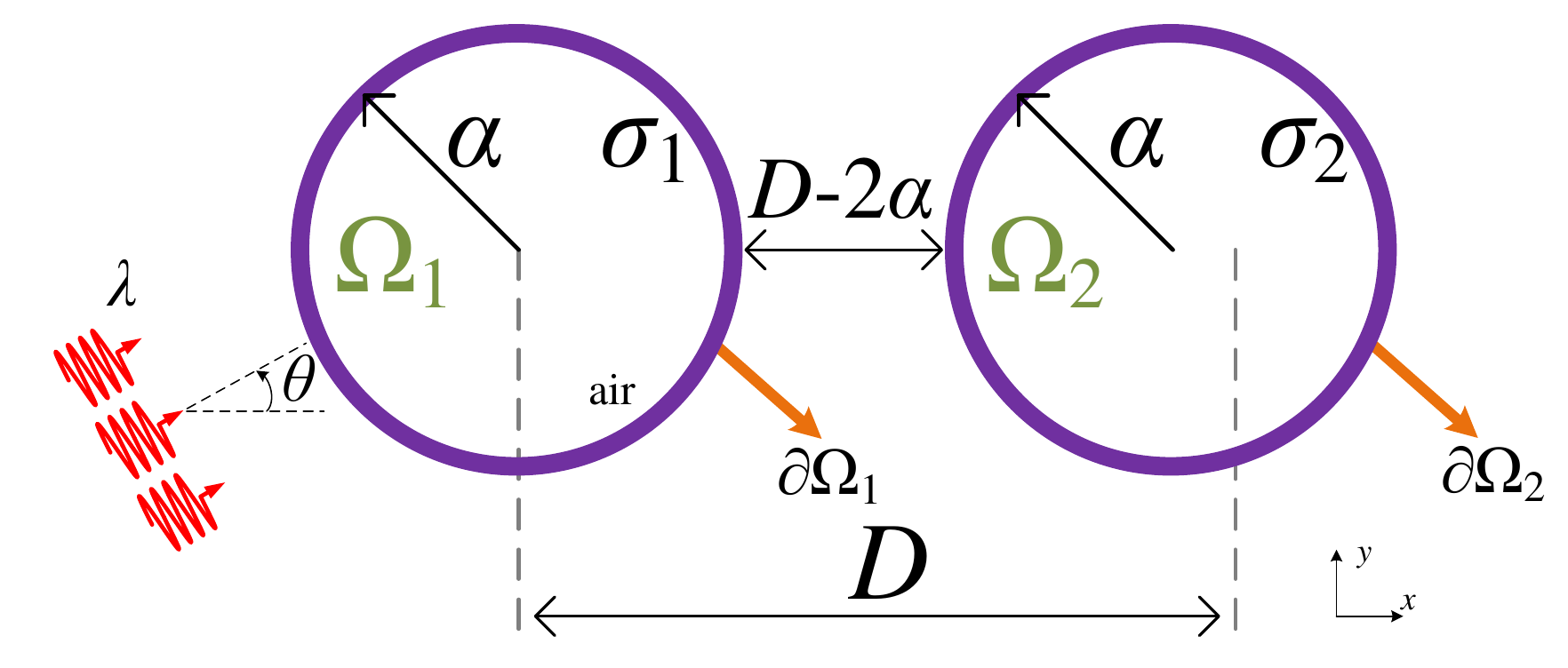}
        \caption{}
        \label{fig:pair_of_circles}
    \end{subfigure}
    \hfill
    \begin{subfigure}[t]{0.49\textwidth}
        \centering
        \includegraphics[width=\textwidth]{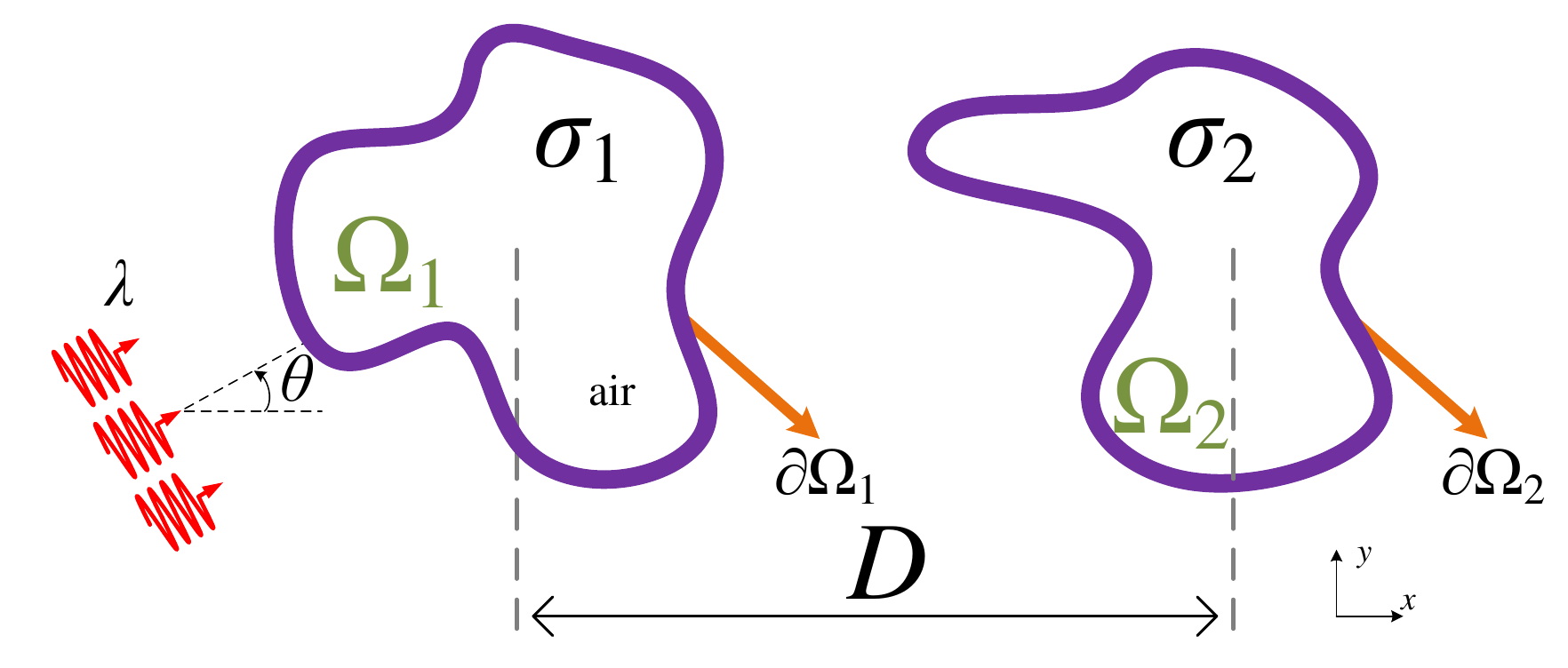}
        \caption{}
        \label{fig:pair_of_freeform_tubes}
    \end{subfigure}
    \caption{\protect Nanotubes excited by a planar TM wave. (a) Pair of circular nanotubes with area $A=|\W_1|=|\W_2|=\pi\alpha^2$, surface conductivities $\sigma_1$, $\sigma_2$, and distance $D$ between their centers. (b) Arbitrary-shaped nanotubes pair with areas $A_1=|\W_1|$, $A_2=|\W_2|$, surface conductivities $\sigma_1$, $\sigma_2$, and distance $D$ between their area centroids.}
    \label{fig:problem_sketch}
\end{figure}

\section{Problem Formulation and Methodology}\label{problem formulation}

\subsection{Electromagnetic Scattering Integral}\label{sec:EMSI}
We consider a planar electromagnetic wave with transverse magnetic (TM) polarization and wavelength $\lambda$ illuminating a pair of circular (Fig.~\ref{fig:pair_of_circles}) or arbitrary-shaped (Fig.~\ref{fig:pair_of_freeform_tubes}) nanotubes. The magnetic field will be on the $xy$-plane with the electric field being normal to the plane, namely parallel to the $z$-axis with reference to Fig.~\ref{fig:problem_sketch}. In the general case, the illuminated nanotubes have different arbitrary shapes, defining two separate domains $\W_1,\W_2$ with varying areas, i.e., $A_1=|\W_1|,\,A_2=|\W_2|$, and different complex surface conductivities $\sigma_1$ and $\sigma_2$, respectively; see also Fig.~\ref{fig:pair_of_freeform_tubes}.

The total electric field, $E(\mathbf{P})$, within the aforementioned domains, as well as outside the nanotubes, equals to the background field, $E_{back}(\mathbf{P})$, plus any scattering created from the nanotube boundaries, $E_{scat}(\mathbf{P})$, with $\mathbf{P}$ corresponding to a point on the $xy$-plane. Following the derivations in~\cite{ChenToTai,MyOldPaper,MyOldPaper2,MyOldPaper3,KOSTAS2023}, we may express the scattering term as
\begin{equation}\label{eq:scattering_term}
E_{scat}(\mathbf{P}) = -\ii k_0 \eta_0 \sum_{i=1}^{2}\oint_{\partial\W_i}G(\mathbf{P},\mathbf{Q})K_i(\mathbf{Q})d\ell_Q
\end{equation}
where $k_0=2\pi/\lambda$ is the wavenumber, $\eta_0=120\pi$ Ohm is the wave impedance into vacuum, $K_i$ denote the axial surface currents which are proportional to the nanotube's surface conductivity, $\sigma_i$, and finally, $G=-\frac{\ii}{4}H_0^{(2)}(k_0\|\mathbf{P}-\mathbf{Q}\|)$ is the corresponding Green's function with $H_0^{(2)}$ denoting the zeroth order Hankel function of the second kind. More specifically, for the axial currents, we have: $\mathbf{K}_i = K_i\mathbf{z}=-\sigma_i\mathbf{n}_i\times(\mathbf{n}_i\times\mathbf{E}_{i})$ where $\mathbf{z}$ is the unit vector in the $z$-direction, $\mathbf{E}_{i}=\mathbf{z} E_{i}$ are the axial electric fields at each of the nanotubes and $\mathbf{n}_i,\,i=1,2$ are the unit normal vectors for the two boundaries. In this way,  \eqref{eq:scattering_term} is written as: $E_{scat}(\mathbf{P}) = -\ii k_0 \eta_0 \sum_{i=1}^2  \sigma_i \oint_{\partial\W_i}G(\mathbf{P},\mathbf{Q})E_{i}(\mathbf{Q})d\ell_Q$. 

Finally, if we impose the conditions for the continuity of the electric field, namely, that: $E_{i}(\mathbf{P}) = E_{back}(\mathbf{P}) + E_{scat}(\mathbf{P})$ at $\mathbf{P}\in\partial\W_i$ for $i=1,2$, we reach the following system of equations 
\begin{equation}\label{eq:BIE}
E_i(\mathbf{P}) + \ii k_0 \eta_0\sum_{i=1}^2\sigma_i \oint_{\partial\W_i}G(\mathbf{P},\mathbf{Q})E_{i}(\mathbf{Q})d\ell_Q  = E_{back}(\mathbf{P}),\quad i=1,2,
\end{equation}
that can be solved with respect to $E_1(\mathbf{Q}), E_2(\mathbf{Q})$ at the nanotube boundaries. Note that, in all considered examples, the background electric field reads: $E_{back}(\mathbf{P})=E_0\exp(-\ii k_0 (x \cos\theta+y \sin\theta))$. 

\subsection{Isogeometric Analysis Boundary Element Method}\label{sec:IGABEM}
The IsoGeometric Analysis Boundary Element Method (IGABEM) approach in this case pertains to the expression of the unknown physical quantity, i.e., the electric fields on the two nanotubes' boundaries, with the same function space used in the representation of their geometry, i.e., $\partial\W_1$ and $\partial\W_2$; see also Fig.~\ref{fig:problem_sketch}. We select to represent these boundaries with two simple closed planar NURBS curves, $\bm{r}_1(u^{(1)}),\, u^{(1)}\in I_1\subset\mathbb{R}$ and $\bm{r}_2(u^{(2)}),\, u^{(2)}\in I_2\subset\mathbb{R}$. If we further assume two appropriate partitions, the so-called knot vectors, $\mathcal{I}_1,\mathcal{I}_2$ of $I_1,I_2$ defined as $\mathcal{I}_1=\{u^{(1)}_0,u^{(1)}_1,\ldots,u^{(1)}_{n_1+k_1}\}$ and $\mathcal{I}_2=\{u^{(2)}_0,u^{(2)}_1,\ldots,u^{(2)}_{n_2+k_2}\}$, we may write the two curves as
\begin{equation}\label{eq:nurbs_curves}
    \bm{r}_i(u^{(i)}) = \sum_{j=0}^{n_i}\mathbf{b}_{i,j}R_{j,k_i}(u^{(i)}),\quad u^{(i)}\in[u^{(i)}_{k_i-1},u^{(i)}_{n_i-1}],\quad i=1,2.
\end{equation}
With respect to the symbols appearing in  \eqref{eq:nurbs_curves}, $k_i$ is the order (degree+1) of the corresponding curve, whereas $\{R_{j,k}(t)\}_{j=0}^{n}$ are the rational basis functions defined over the corresponding knot vector, i.e., $R_{j,k}=\frac{w_j N_{j,k}(t)}{\sum_{\ell=1}^{n}w_\ell N_{\ell,k}(t)}$, with $N_{j,k}$ denoting the $j$-th $k$-order B-Spline basis function defined over the same knot vector; see also~\cite{PieglTiller97} for more details on B-Splines and NURBS curves. 

Therefore, in the context of the IGABEM approach, the same rational basis functions should be used to represent the unknown quantities $E_1$ and $E_2$, namely, $E_i(u^{(i)}) = \sum_{j=0}^{n_i}e_{i,j}R_{j,k_i}(u^{(i)})$, where $u^{(i)}\in[u^{(i)}_{k_i-1},u^{(i)}_{n_i-1}]$ and $i=1,2$. Since the number of basis functions, $n_1+1$ and $n_2+1$, may not suffice for a good approximation of $E_1$ and $E_2$, we consider enriched spline spaces with $m_1, m_2 \in\mathbb{N}^*$ additional knots being inserted in $\mathcal{I}_1$ and $\mathcal{I}_2$, respectively. This leads to the following, more general representation of the electric field on the two nanotubes:
\begin{equation}\label{eq:e_curves2}
        E_i(u) = \sum_{j=0}^{n_i+m_i}e_{i,j}R_{j,k_i}^{(m_i)}(u^{(i)}),\quad u^{(i)}\in[u^{(i)}_{k_1-1},u^{(i)}_{n_1-1}],\\
\end{equation}
where $R_{j,k}^{(0)}(t)\equiv R_{j,k}(t)$. If we now plug  \eqref{eq:e_curves2} into  \eqref{eq:BIE} we get the following system of two integral equations ($i=1,2$):
\begin{equation}\label{eq:BIE-IGA}
    \begin{aligned}
     \sum_{j=0}^{n_i+m_i}e_{i,j}R_{j,k_i}^{(m_i)}(u^{(i)}) + \ii k_0 \eta_0\sum_{\ell=1}^2 \sigma_\ell \sum_{j=0}^{n_\ell+m_\ell}e_{\ell,j}\int_{I_\ell}G(\bm{r}_i(u^{(i)}),\bm{r}_\ell(t))\|\dot{\bm{r}}_\ell(t)\|dt \\ = E_{back}(\bm{r}_i(u^{(i)})).\\
    \end{aligned}
\end{equation}

Similar to the implementation of the IGABEM approach described in~\cite{KOSTAS2023}, we employ here a collocation method to solve the system in  \eqref{eq:BIE-IGA}. For the collocation points, Greville abscissae~\cite{PieglTiller97,greville} from the corresponding knot vectors ($\mathcal{I}_1^{(m_1)}$ and $\mathcal{I}_2^{(m_2)})$ are used, thereby resulting in a combined linear system with $(n_1+m_1+1)+(n_2+m_2+1)$ equations and a relatively low condition number. Furthermore, a problem-informed refinement procedure permits the achievement of high convergence rates as has been demonstrated in~\cite{Kostas2024}. Greville abscissae are defined as knot averages, i.e., $\hat{t}_i=(t_{i+1}+t_{i+2}+\ldots+t_{i+k-1})/(k-1)$, and hence, the resulting linear system becomes

\begin{equation}\label{eq:BIE-IGA-final}
    \begin{aligned}
     \sum_{j=0}^{n_i+m_i}e_{i,j}R_{j,k_i}^{(m_i)}(\hat{u}_{{\alpha_1+\alpha_2}i}^{(i)}) + \ii k_0 \eta_0\sum_{\ell=1}^2 \sigma_\ell \sum_{j=0}^{n_\ell+m_\ell}e_{\ell,j}\int_{I_\ell}G(\bm{r}_i(\hat{u}_{\alpha_i}^{(i)}),\bm{r}_\ell(t))\|\dot{\bm{r}}_\ell(t)\|dt \\ = E_{back}(\bm{r}_i(\hat{u}_{\alpha_i}^{(i)})),\quad a_i = 0,1,\ldots,n_i+m_i, \quad i=1,2.\\
    \end{aligned}
\end{equation}

Subsequently, using the solution of  \eqref{eq:BIE-IGA-final}, i.e., $\{e_{1,j}\}_{j=0}^{n_1+m_1}$ and $\{e_{2,j}\}_{j=0}^{n_2+m_2}$, we can determine the respective electric fields, $E_1,\, E_2$, on the two boundaries, $\partial\W_1$, $\partial\W_2$, using  \eqref{eq:e_curves2}. Finally, we may plug $E_1(\mathbf{Q}),\,E_2(\mathbf{Q})$ into  \eqref{eq:BIE} and compute the value of $E_1(\mathbf{P})$, $E_2(\mathbf{P})$, with $\mathbf{P}\in\W_1\cup\W_2$. Therefore, the electric field concentration, in the interior of the two nanotubes, becomes:
\begin{equation}\label{eq:concentration}
    Q = Q_1+Q_2 = \int_{\W_1}E_1(\mathbf{P})ds + \int_{\W_2}E_2(\mathbf{P})ds. 
\end{equation}
As a final note, it is worth mentioning here that the IGABEM approach described above has been successfully validated for the case of a single, circular or of arbitrary shape, nanotube against analytic and numerical solutions while achieving very high convergence rates; see~\cite{KOSTAS2023,IZNAT2025106153} for more details.

\subsection{Coupled Circular Nanotubes}\label{sec:parametric_study}
If we assume two circular nanotubes with their centers on the $x-$axis as shown in Fig.~\ref{fig:problem_sketch}, the value of the electric field concentration, $Q$ from  \eqref{eq:concentration}, will clearly depend on the following quantities:
\begin{itemize}
    \item $A_i/\lambda^2$: ratios of nanotube's area over the squared wavelength ($i=1,2$);
    \item $\Re(\sigma_i\eta_0)$ and $\Im(\sigma_i\eta_0)$: the complex surface electric conductivity of the employed nanotube materials ($i=1,2$);
    \item $D/(\alpha_1+\alpha_2)$: the distance between the nanotubes' centers;
    \item $\theta$: the angle of incidence of the EM wave as defined in Fig.~\ref{fig:problem_sketch}.
\end{itemize}
Hence, we may generally write:
$Q = \sum_{i=1}^2Q_i\left(\frac{A_1}{\lambda^2},\frac{A_2}{\lambda^2},\sigma_1\eta_0,\sigma_2\eta_0,\frac{D}{\alpha_1+\alpha_2},\theta\right),
$
where obviously $A_i=\pi\alpha_i^2$. In this work, we reduce this high-dimensional space by assuming the nanotubes are of the same material, equal size, and equidistant to the origin, hence $\alpha_1=\alpha_2=\alpha$, $\sigma_1=\sigma_2=\sigma$, and $D_1=D_2=D/2$, which subsequently leads to: $Q(\bm{\varrho}) = \sum_{i=1}^2Q_i(\bm{\varrho}),\quad\bm{\varrho}=\left(A/\lambda^2,\sigma\eta_0,D/(2a),\theta\right).$
We can further denote as $Q_c$ the concentration of the electric field in a single circular nanotube under the same conditions, which finally yields the normalized concentration
\begin{equation}\label{eq:concentration_norm}
\widetilde{Q}(\bm{\varrho}) = \frac{\sum_{i=1}^2Q_i(\bm{\varrho})}{2Q_c(\bm{\varrho})}. 
\end{equation}
Note that the ratio in \eqref{eq:concentration_norm} equals one once the two circles are positioned at an infinite distance apart ($D\rightarrow +\infty$). In such a scenario, there is no interference between the two identical nanotubes and their response is just two times the value of an isolated one, i.e., the numerator equals the denominator.

\begin{figure}[thbp]
 \centering
  \subfloat[$\widetilde{Q}$ on $(\theta,D/(2\alpha))$ map with $\sigma\eta_0=0+\ii 5.3$. \label{fig:p_study_e1}]{\includegraphics[width = 0.45\textwidth]{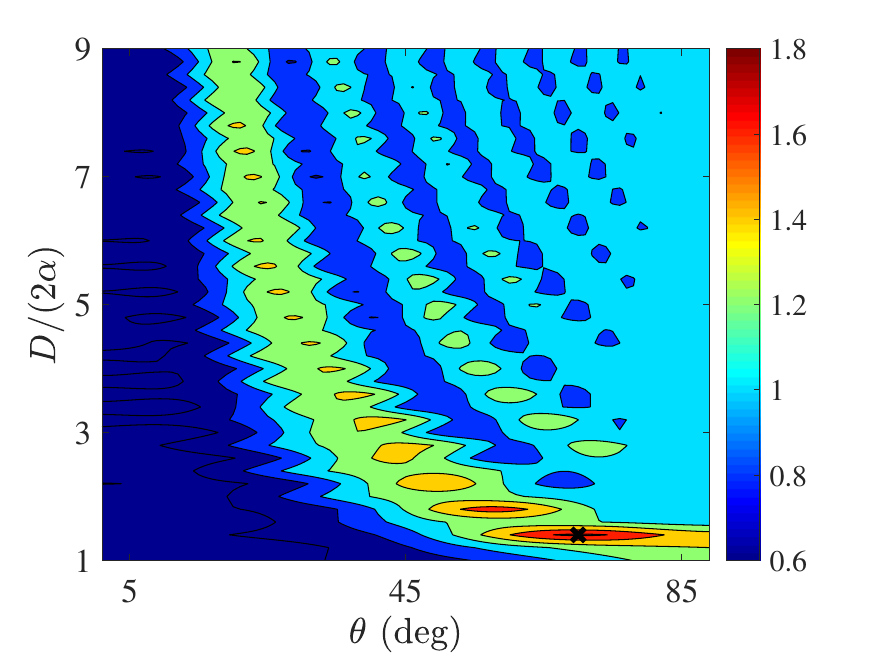}}\quad
  \subfloat[$\max_{\theta}(\widetilde{Q})$ on
$(\Im(\sigma\eta_0),D/(2\alpha))$ map.\label{fig:p_study_e2}]{\includegraphics[width = 0.45\textwidth]{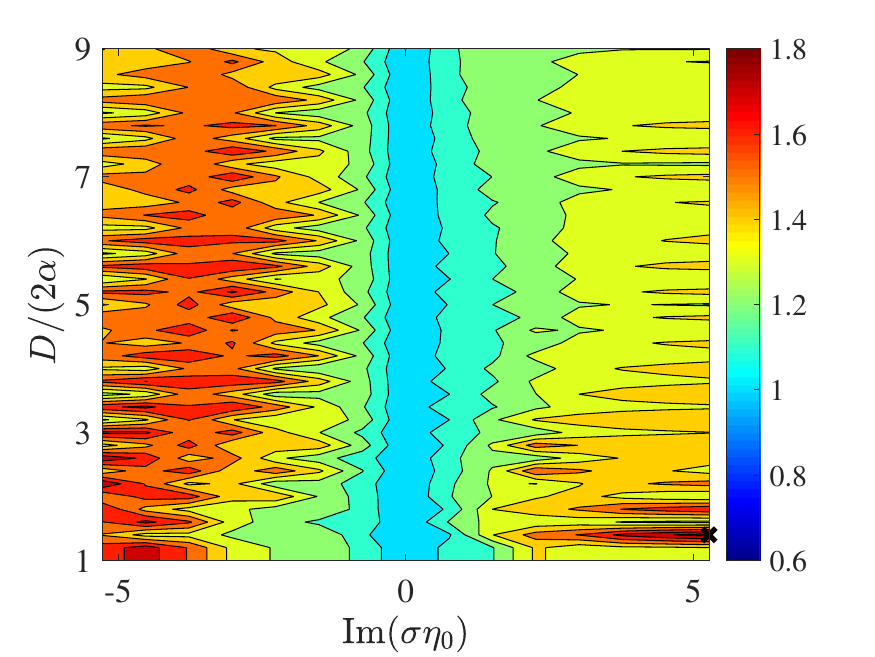}}
  \caption{Slices from the parametric space for $A/\lambda^2=1$. (a) Variation of $\widetilde{Q}$ with respect to $\theta$ and $D/(2\alpha)$. (b) Variation of $\max_{\theta}(\widetilde{Q})$ with respect to $\Im(\sigma\eta_0)$ and $D/(2\alpha)$. Note that the position of maximum $\widetilde{Q}$ is marked with a black $\bm{\times}$ marker on both plots. Lossless ($\Re(\sigma\eta_0)=0$) designs are considered.}
  \label{fig:p_study_example}
\end{figure}

The abovementioned design-space restriction leads to a 5-dimensional parametric space (recall that $\sigma$ is a complex number), which is systematically studied in this work. Specifically, we limited the values of the involved quantities (see Table~\ref{tab:parameters}) to derive 2D slices of the design space as shown, for example, in Fig.~\ref{fig:p_study_example} for $A=\lambda^2 \Rightarrow \alpha=\lambda/\sqrt{\pi}$ and $\sigma\eta_0= 0+5.3\ii$. Indeed, a common choice is to consider lossless designs ($\Re(\sigma\eta_0)=0$) since we care  about field concentration; absorption usually subtracts from it. In particular, in Fig.~\ref{fig:p_study_e1}, we represent the basic metric $\widetilde{Q}$ from \eqref{eq:concentration_norm} with respect to the incidence angle $\theta$ and the normalized distance between the centers of the two lossless nanotubes $D/(2\alpha)$.  One typically observes multiple peaks that resemble Fabry-Perot distance resonances~\cite{10314452} appearing along hyperbolic-shaped paths on the considered plane. As anticipated, the maxima get weaker as the circles get farther apart and, inevitably, $\widetilde{Q}\rightarrow 1$ for $D\rightarrow +\infty$. The setup with the highest performance is denoted by a black $\bm{\times}$ marker in Fig.~\ref{fig:p_study_e1}. We should also note here that we only need to examine values of the $\theta$ angle between $0$ and $90^\circ$ since, due to the symmetry about the $x-$axis, the same results would be obtained for values between $0$ and $-90^\circ$. Finally, for values between $90$ and $270=-90^\circ$ the results would be only swapped between $Q_1$ and $Q_2$ leading to the same total value of $Q$.

As one may easily see, if we were to present the full design space using slices as in Fig.~\ref{fig:p_study_e1}, an overwhelmingly large number of such parametric slices would be needed. For that reason, in Fig.~\ref{fig:p_study_e2}, we suppress the angle parameter $\theta$  by only assuming illumination directions at which $\widetilde{Q}$ takes its maximum value. Specifically, Fig.~\ref{fig:p_study_e2} depicts the quantity $\max_\theta(\widetilde{Q})$ as a function of the imaginary part of the normalized electrical conductivity $\Im(\sigma\eta_0)$ and the separating distance $D/(2\alpha)$. Across the zone of small $|\Im(\sigma\eta_0)|$, the signal power into the cylinders equals to that of the background field ($\widetilde{Q}=1$), since nanotubes become transparent therein. It is noteworthy that plasmonic cylinders ($\Im(\sigma\eta_0)<0$) collect  more efficiently the incident beam compared to their dielectric ($\Im(\sigma\eta_0)>0$) counterparts; see Fig.~\ref{fig:p_study_e2}. However, the most successful design for this slice possesses a positive $\Im(\sigma\eta_0)=5.3$, as indicated by the black $\bm{\times}$ marker. 

The full parametric space can be presented in only 20 plots shown in the Appendix; see Figs.~\ref{fig:p_study_1} to~\ref{fig:p_study_4}. One may easily observe from the same set of figures, i.e., Figs.~\ref{fig:p_study_1} to~\ref{fig:p_study_4}, that circular nanotube pairs show, in most cases, relatively elevated field concentrations when $1+\epsilon\leq D/(2\alpha)\leq3.5$ and $|\Im(\sigma\eta_0)|\geq2$. In addition, the best performing circular nanotube pairs achieve an improvement between 1.8 to 2.65 times the field concentration of two circular nanotubes at infinite distance; see again Figs.~\ref{fig:p_study_1} to~\ref{fig:p_study_4}. It is also noteworthy that large regions of the parametric space yield poor results, i.e., $\widetilde{Q}<1$; see, for example, the blueish regions in Fig.~\ref{fig:p_study_e1}, which indicates the challenge in achieving a substantially constructive coupling between two circular nanotubes.

\begin{table}[htb]
\centering
\begin{tabular}{l  c c l}%% Table column specifiers
\hline
Parameter &  Lower limit & Upper limit \\
\hline\hline
$A/\lambda^2$ & 0.5 & 10 \\ \hline
$\Re(\sigma\eta_0)$ & 0 & 5.3\\ \hline
$\Im(\sigma\eta_0)$ & -5.3 & 5.3 \\ \hline
$D/(2a)$ & $1+\epsilon$ & 10 \\ \hline
$\theta$ & $0^\circ$ & $90^\circ$ \\ \hline
\end{tabular}
\caption{Limits of parameters which are being swept to produce the graphs of the Appendix for two coupled circular nanotubes. The symbol $\epsilon$ is used for a small positive number: ($0<\epsilon\ll 1$).}
\label{tab:parameters}
\end{table}

\subsection{Field Concentration Maximization}\label{sec:optimization}
The main objective of this work is to determine the shapes of two nanotubes which maximize the value of $Q$ (see  \eqref{eq:concentration}) over the corresponding value of the same quantity for a pair of circular nanotubes, under the same EM wave source and relative layouts. For formulating this optimization problem, we consider three domains, $\W_1$, $\W_2$, and $\W_c$, along with the corresponding domain boundaries, $\partial\W_1$, $\partial\W_2$, and $\partial\W_c$, which correspond to two arbitrary-shaped and a circular nanotube, respectively. For the selection of the circular nanotube and the problem layout, we consider one point of interest in the regions considered in the parametric study; see \S\ref{sec:parametric_study}. As soon as such a point is selected we have also determined the values of $A/\lambda^2=|\W_c|/\lambda^2$, $\sigma\eta_0$, $D/(2\alpha)$, and $\theta$. If we additionally consider $\mathbf{v}\in\mathcal{V}\subset\mathbb{R}^n$ to be the vector of $n$ design variables that determine the shape of each nanotube, i.e., $\partial\W_1(\mathbf{v}_1)$ and $\partial\W_2(\mathbf{v}_2)$, we may formulate the shape optimization problem as follows:
\begin{equation}\label{eq:shape_optmization}
    \begin{aligned}
        \mathrm{find}\,\mathbf{v}_1^*,\mathbf{v}_2^* : &\quad Q'(\mathbf{v}_1^*,\mathbf{v}_2^*) = \underset{\mathbf{v}_1,\mathbf{v}_2}{\max} \frac{Q_1(\mathbf{v}_1)+Q_2(\mathbf{v}_2)}{Q_{c1}(\partial\W_{c1})+Q_{c2}(\partial\W_{c2})} \\
        \mathrm{subject\,to} : &\quad\mathbf{v}_1,\mathbf{v}_2\in\mathcal{V},\\
    &\quad\partial\W_1(\mathbf{v}_1),\partial\W_2(\mathbf{v}_2)\,\mathrm{are\,Jordan\,curves},\\
                               &\quad\W_1(\mathbf{v}_1)\cap\W_2(\mathbf{v}_2) = \varnothing,\\ &\quad|\W_1(\mathbf{v}_1)|=|\W_2(\mathbf{v}_2)|=|\W_{c1}|=|\W_{c2}|=\pi\alpha^2,\\
                               &\quad \mathbf{c}_1(\mathbf{v}_1)=(-D/2,0),\,\mathbf{c}_2(\mathbf{v}_2)=(D/2,0),\\
                               &\quad \mathrm{and}\,\lambda,\sigma,\theta\, \mathrm{are\,given},
    \end{aligned}
\end{equation}
where $Q_i(\mathbf{v}_i)\equiv Q_i\left(\partial\W_i(\mathbf{v}_i)\right)$, $\mathbf{c}_i=(c_{ix},c_{iy})$ denotes the centroid of $\W_i$, and the sum of $Q_{c1}(\partial\W_{c1})+Q_{c2}(\partial\W_{c2})$ corresponds to the field concentration of the pair of equiareal circular nanotubes under the same operational conditions. For the solution of the shape optimization problem in  \eqref{eq:shape_optmization}, we need an efficient and robust way of determining the shape of $\partial\W_i$ from a design vector $\mathbf{v}_i$, which is discussed in detail in~\S\ref{sec:pmodels}. At the same time, the two nonlinear constraints, involving each nanotube's area and centroid, are also handled during the construction of $\partial\W_i(\mathbf{v}_i)$, as it will be also discussed in~\S\ref{sec:pmodels}. Obviously, handling the non-linear constraints at the shape-construction phase simplifies significantly the solution of the optimization problem above. Finally, a wide range of local and global optimizers, including \textsc{matlab}\footnote{\url{https://www.mathworks.com/products/matlab.html}}'s fmincon~\cite{10.1007/PL00011391,doi:10.1137/S1052623497325107}, patternsearch~\cite{doi:10.1137/S1052623400378742,doi:10.1137/0728030,doi:10.1137/S003614450242889}, ga~\cite{10.5555/534133,doi:10.1137/0728030}, and PAGMO's~\cite{Biscani2020} cobyla~\cite{Powell1994}, compass search~\cite{doi:10.1137/S003614450242889}, improved harmony search (IHS)~\cite{MAHDAVI20071567}, have been applied for the solution of the shape optimization problem in  \eqref{eq:shape_optmization}. In brief, fmincon is a local, gradient-based, nonlinear programming solver that supports nonlinear constraints. Patternsearch is \textsc{matlab}'s enhanced variation of the classical derivative-free, direct search method also known as compass search, while ga is \textsc{matlab}'s implementation of a single-objective Genetic Algorithm with support of linear and nonlinear constraints. Cobyla is a derivative of Powell's implementation of the `Constrained Optimization by Linear Approximations' algorithm for derivative-free optimization with nonlinear inequality and equality constraints, and finally, IHS is a metaheuristic algorithm said to mimic the improvisation process of musicians. In the metaphor, each musician (i.e., each variable) plays (i.e., generates) a note (i.e., a value) for finding a best harmony (i.e., the global optimum) all together. It belongs to the same family of evolutionary algorithms as \textsc{matlab}'s ga implementation mentioned above. The optimization problems in this work have been addressed by employing a global optimization algorithm (e.g., patternsearch, ga or IHS) to approximate the optimum design, followed by a local optimizer to expedite the search when a sufficiently good design point has been reached.

\subsection{Parametric Models}\label{sec:pmodels}
As already mentioned in \S\ref{sec:IGABEM}, the two boundaries will be represented via two NURBS curves, which also need to be Jordan curves, i.e., planar simple closed curves, to avoid unwanted self-intersections. IGA-based shape optimization via NURBS representations has been extensively studied in pertinent literature, e.g.,~\cite{WallEtAl08,NguyenEtAl2012,LiQian11,Kostas2015,KOSTAS2018}, however, the direct use of control points poses various difficulties in satisfying the problem's requirements which would consequently require a number of complicated constraints that would significantly increase the complexity of the optimization problem at hand. For that reason, two parametric models for the generation of the nanotube's shape, $\partial\W$, have been employed in this work, with the aim of robustly generating the two Jordan curves for any design vector $\mathbf{v}\in\mathcal{V}$ while automatically satisfying the area and positional constraints posed in  \eqref{eq:shape_optmization}.

\begin{figure}[thbp]
 \centering
  \subfloat[standard arbitrary shape\label{fig:pmodel1}]{\includegraphics[width = 0.45\textwidth]{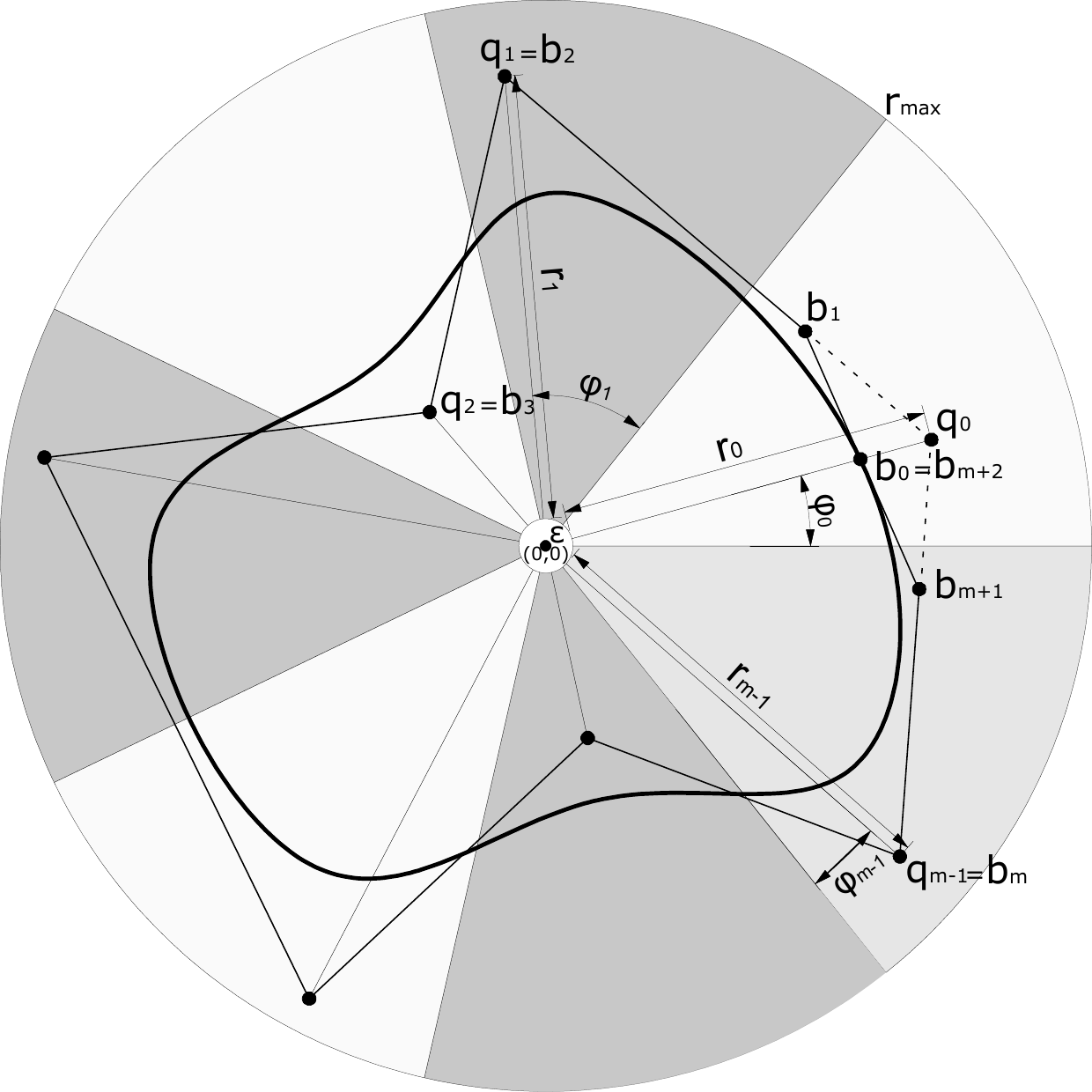}}\quad
  \subfloat[x-symmetric shape\label{fig:pmodel2}]{\includegraphics[width = 0.45\textwidth]{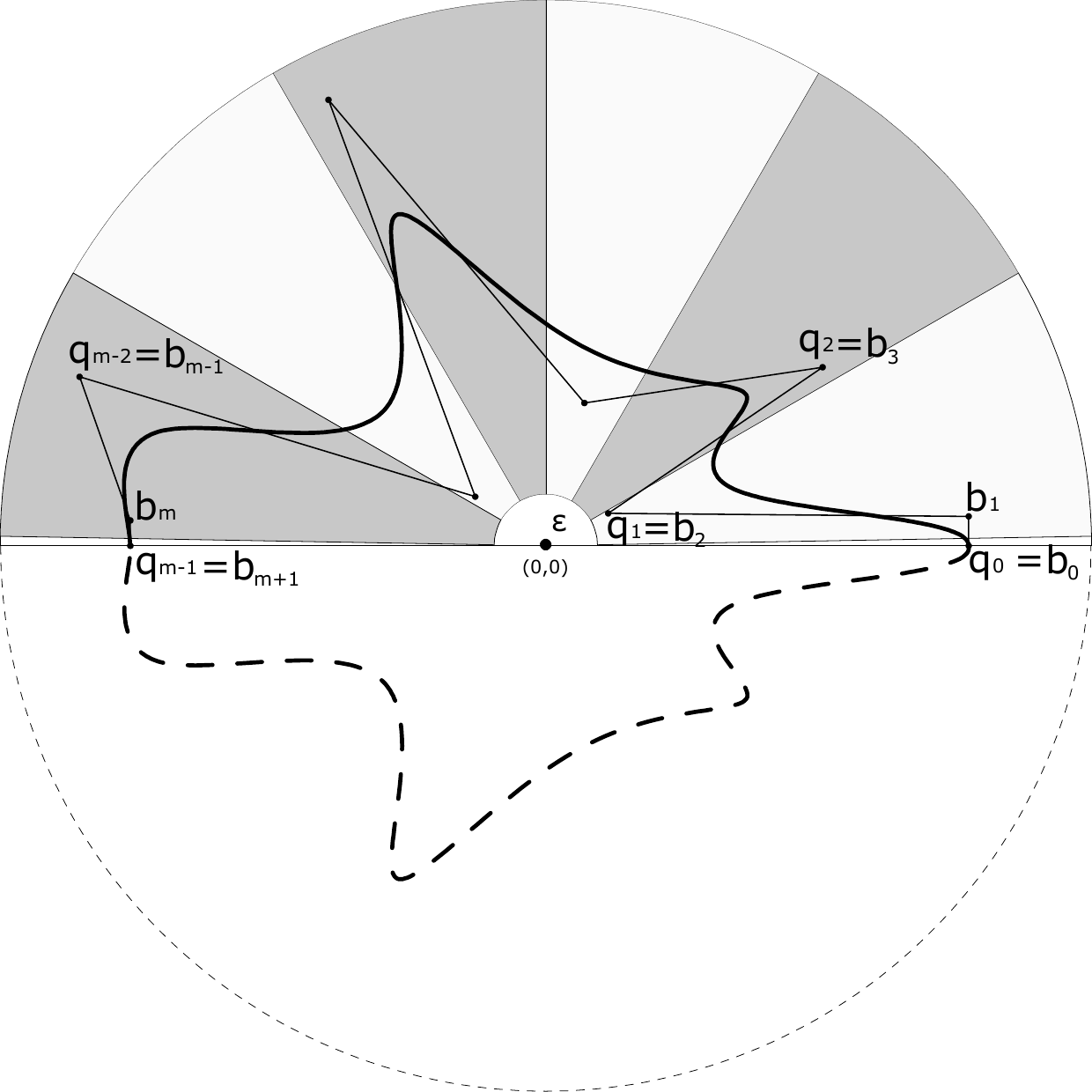}}
  \caption{Parametric models employed in this work.}
  \label{fig:ParametricModels}
\end{figure}

The first parametric model is adopted from~\cite{KOSTAS2023} in which $m$ shape-controlling points, $\{\mathbf{q}\}_{i=0}^{m-1}$, are placed in $m$ equal-sized annulus sectors covering a full circle, as shown in Fig.~\ref{fig:pmodel1}. The placement of these points is determined by $n=2m$ parameters, $\mathbf{v}=(v_0,v_1,\ldots,v_{n-1})$, as follows:
\begin{equation}\label{eq:pmodel1}
\begin{aligned}
    \mathbf{q}_i = & \left[r_i\cos\left(\phi_i+\frac{2\pi i}{m}\right),r_i\sin\left(\phi_i+\frac{2\pi i}{m}\right)\right]^T \\ 
    \mathrm{where}\quad & r_i = v_i(r_{\max}-\epsilon)+\epsilon,\quad\phi_i=\frac{2\pi v_{i+m}}{m},\\
    & \mathbf{v}\in[0,1]^{2m},\,\mathrm{i.e.,}\,v_i,v_{i+m}\in[0,1],\,i=0,1,\ldots,m-1,
\end{aligned}
\end{equation}
and $\epsilon$ denotes an arbitrarily small positive number with $r_{\max}$ being the radius of the circle enclosing the shape. Subsequently, these shape-controlling points along with some additional automatically-constructed control points (see points marked with $\mathbf{b}$ in Fig.~\ref{fig:pmodel1}), which satisfy clamped or periodic boundary conditions, constitute the set of control points defining $\partial\W$. Note here that the value of $r_{\max}$ can be practically set to any positive value as the final closed curve is scaled and translated so that its enclosing domain matches the area and centroid constraints; see  \eqref{eq:shape_optmization}. A more detailed description of this construction can be found in~\cite{KOSTAS2023}.

In addition to the standard parametric model depicted in Fig.~\ref{fig:pmodel1}, another parametric model which constructs a planar simple closed curves with symmetry about the $x-$axis has been developed and depicted in Fig.~\ref{fig:pmodel2}. This new model allows the construction of complicated shapes with a smaller number of parameters while at the same time simplifying the placement of arbitrary shaped so that their centroids are properly positioned. 
More significantly, it avoids the $\theta$-angle-alignment problem that occurs when the first model is used, i.e., the shape-alignment of the nanotube to the angle of the incoming EM wave; see the relevant discussion in \S\ref{sec:results}.

For the construction of this second parametric model, the following modifications have been implemented:
\begin{itemize}
    \item the first and last shape-controlling points, $\mathbf{q}_0$ and $\mathbf{q}_{m-1}$, are placed at $(r_0,0)$ and $(r_{m-1},\pi)$, respectively;
    \item two additional points, $\mathbf{b}_1$ and $\mathbf{b}_m$, are constructed on the vertical lines passing through $\mathbf{q}_0$ and $\mathbf{q}_{m-1}$, respectively. These points secure the tangential continuity between the two parts. Furthermore, their exact positions are determined by $v_m$ and $v_{n-1}$, which are percentages of the neighboring point ordinates to avoid potential self-intersections in that region; 
    \item the $m$ sectors are distributed over the half disk, i.e., from $0$ to $\pi$, with the first and last annulus sectors being slightly reduced (as can be seen in Fig.~\ref{fig:pmodel2}) to avoid $\mathbf{q}_0$ and $\mathbf{q}_1$ (and similarly $\mathbf{q}_{m-2}$ and $\mathbf{q}_{m-1}$) both lying on the $x-$axis. 
    \item Finally, the constructed upper part is mirrored with respect to the $x-$axis and a single NURBS curve is constructed by the two parts.
\end{itemize}

Both the standard and $x-$symmetric models automatically satisfy the area and centroid constraints posed by the optimization problem, along with the non-self-intersecting requirement. Specifically, the area constraint is satisfied by scaling the resulting model's instance accordingly whereas an appropriate shape translation is sufficient to place the constructed shape's centroid on the required position.  However, an additional step is required to secure against possible intersections between the two curves. This is accomplished by performing a quick curve-curve intersection algorithm implementation using polygonal approximations and automatically penalizing the design vectors that result in intersecting nanotube pairs.

\section{Numerical Results}\label{sec:results}

As described in section ~\ref{sec:parametric_study}, the parametric study was conducted to examine the behavior of circular nanotube pairs with respect to individual nanotubes, and subsequently identify regions of interest which will allow the determination of the 5 required  parameter values for shape optimization, i.e., normalized area ($A/\lambda^2)$, real and imaginary parts of surface conductivity ($\Re(\sigma\eta_0),\Im(\sigma\eta_0)$), the distance between nanotubes' centroids ($D/(2\alpha)$), and the angle between the EM wave and the $x$-axis ($\theta$). Generally, we aim in identifying regions in which the circular nanotubes pair is performing well, i.e., the interaction of the two nanotubes increases the field concentration with respect to two circular nanotubes at an infinite distance to each other. We avoid examining cases with low performance of the circular nanotubes pair, as it would be relatively easy to obtain improvements via shape modification there.

\begin{figure}[htbp]
 \centering
  \subfloat[$\widetilde{Q}(D/(2\alpha),\theta)$ for lossy $\sigma\eta_0=0.754+5.278\ii$ and moderate $A/\lambda^2=1$.\label{fig:Area1PointColormap}]{\includegraphics[width = 0.48\textwidth]{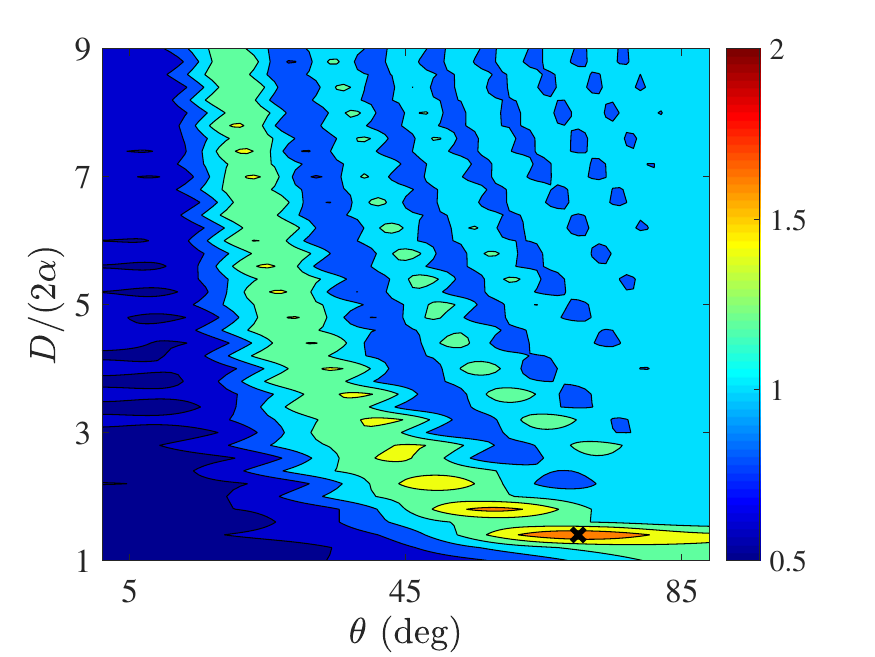}}\hfill
  \subfloat[$\widetilde{Q}(D/(2\alpha),\theta)$ for lossless $\sigma\eta_0=0-2.262\ii$ and sizeable $A/\lambda^2=2.5$.\label{fig:Area2.5PointColormap}]{\includegraphics[width = 0.48\textwidth]{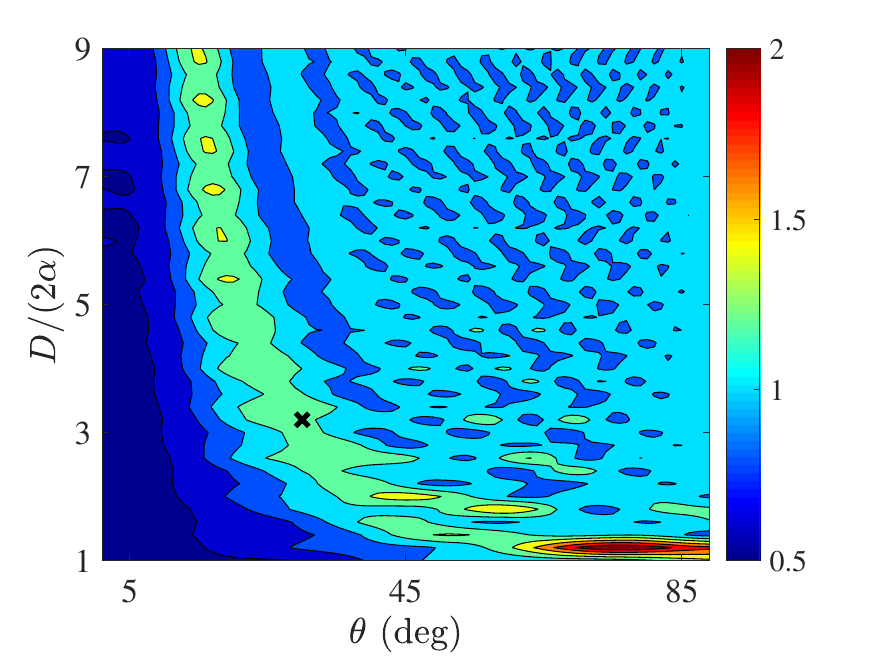}}
  \caption{Two representative cases of coupled circular nanotubes that are selected for shape optimization denoted by black $\bm{\times}$ markers in the design space.}
  \label{fig:PointsColormapOpt}
\end{figure}

In Fig.~\ref{fig:Area1PointColormap}, we show the normalized field concentration $\widetilde{Q}$ across the usual map $(\theta, D/(2\alpha))$ (like that of Fig.~\ref{fig:p_study_e1}) for two representative cases. The first one (Fig.~\ref{fig:Area1PointColormap}) comprises two moderately-sized nanotubes ($A=\lambda^2$) made of a lossy dielectric medium ($\sigma\eta_0\cong 0.754+\ii 5.278$). As far as the second is concerned (Fig.~\ref{fig:Area2.5PointColormap}), the cylinders are optically bigger ($A=2.5\lambda^2$) and the medium lossless and deeply plasmonic.

We proceed by presenting some indicative shape optimization results (see the formulation of the shape optimization problem in  \eqref{eq:shape_optmization}) starting from the two aforementioned designs; the operational conditions are indicated by the black $\bm{\times}$ markers in Fig.~\ref{fig:PointsColormapOpt} and summarized in Table~\ref{tab:opt conditions}. In all examples, we begin by employing a small number of shape-controlling points, SCPs, for the parametric models in \S\ref{sec:pmodels}, and increase their number till no further improvements can be achieved or when these improvements become negligible. We demonstrate the use of the symmetric parametric model in the first case as it can produce sufficiently complex shapes with a lower number of SCPs and does not exhibit the potential problems with wave angle alignment that occur when the non-symmetric model is used; see second case. 

\begin{table}[htb]
\centering
\begin{tabular}{l c c }%% Table column specifiers
\hline
Parameter &  Case I (Fig.~\ref{fig:p1_Results}) & Case II (Fig.~\ref{fig:p2_Results}) \\
\hline\hline
$A/\lambda^2$ & 1  & 2.5  \\ \hline
$\Re(\sigma\eta_0)$  & 0.754  & 0  \\ \hline
$\Im(\sigma\eta_0)$  & 5.278  & -2.262  \\ \hline
$D/(2\alpha)$  & 1.4  & 3.2 \\ \hline
$\theta$ & $70^{\degree}$ &  $30^{\degree}$  \\ \hline
$\widetilde{Q}$ & 1.78 & 1.25  \\ \hline
$Q_{c1}+Q_{c2}$ & 0.292  & 0.959  \\ \hline
\end{tabular}
\caption{Operational conditions of the two representative cases that are selected to enter shape optimization.}\label{tab:opt conditions}
\end{table}

\begin{figure}[htbp]
 \centering
  \subfloat[3 SCPs, $Q' = 14.24$.\label{fig:p1_SCP3}]{\includegraphics[width = 0.45\textwidth]{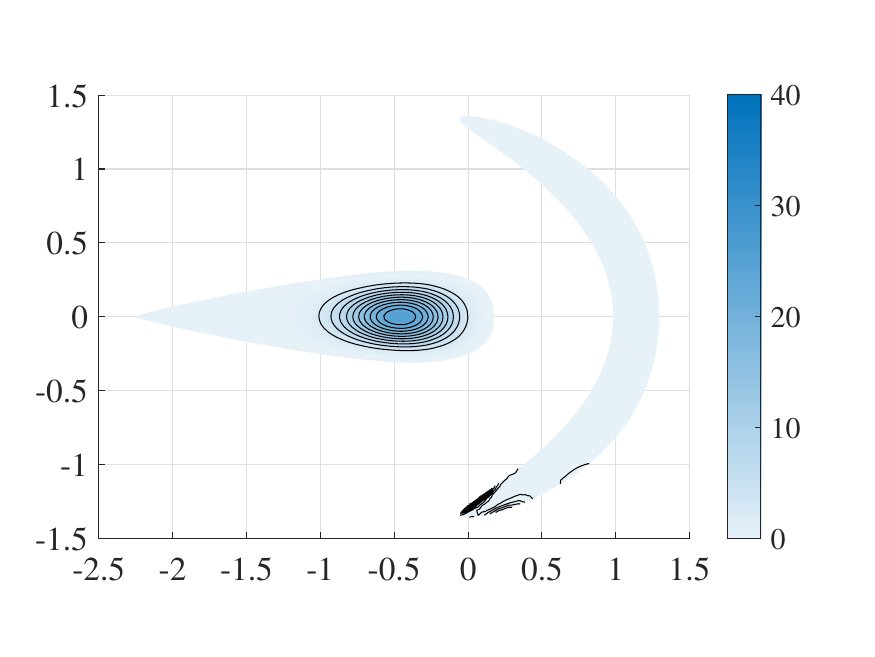}} \hfill
  \subfloat[4 SCPs, $Q' = 17.65$.\label{fig:p1_SCP4}]{\includegraphics[width = 0.45\textwidth]{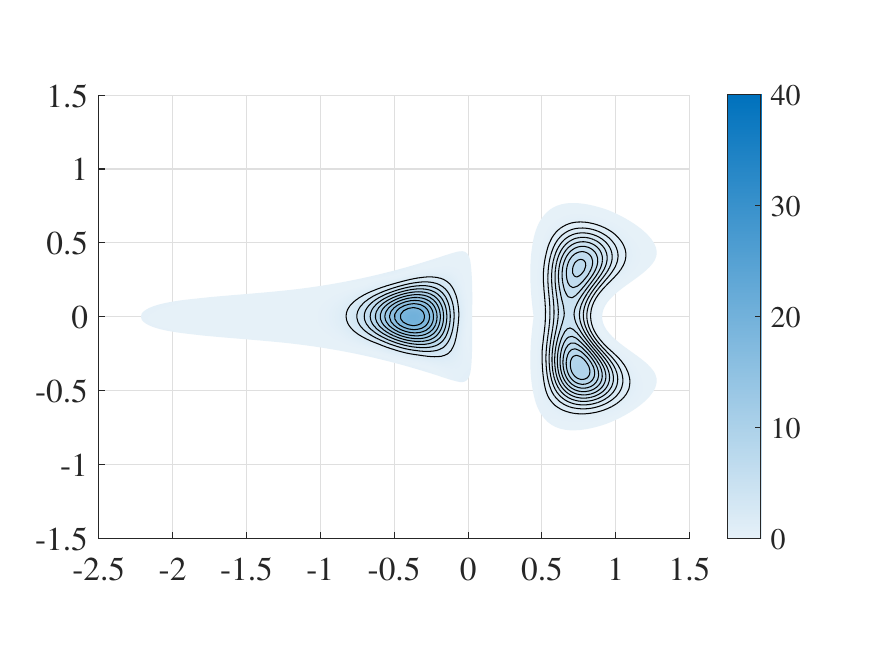}} \\
  \subfloat[5 SCPs, $Q' = 25.17$.\label{fig:p1_SCP5}]{\includegraphics[width = 0.45\textwidth]{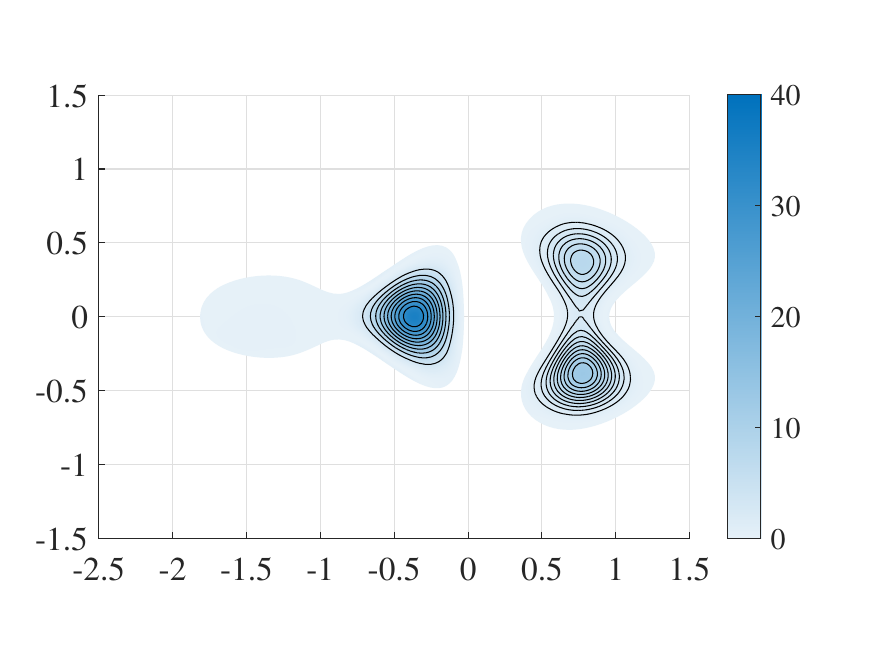}} \hfill
  \subfloat[6 SCPs, $Q' = 32.49$.\label{fig:p1_SCP6}]{\includegraphics[width = 0.45\textwidth]{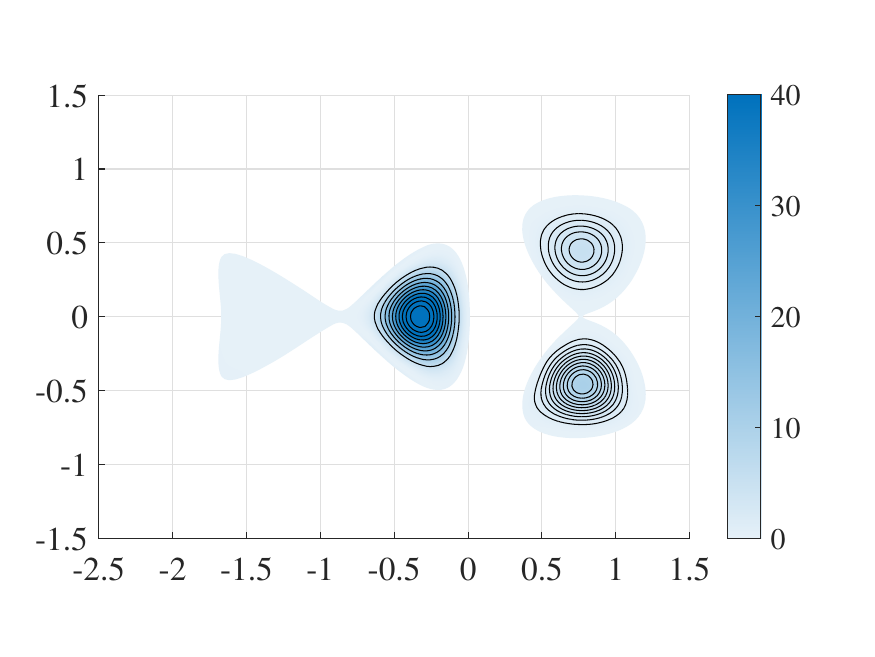}}
  \caption{Optimization results for the Case I with an increasing number of SCPs (3,4,5, and 6) at \mbox{$A/\lambda^2=1$}, $D/(2\alpha)=1.4$, $\theta=70{\degree}$, and $\sigma\eta_0=0.754+5.278\ii$, using the symmetric parametric model.}
  \label{fig:p1_Results}
\end{figure}

The first case (Case I in Table~\ref{tab:opt conditions}) corresponds to the maximum $\widetilde{Q}$ value that can be achieved by a pair of circular nanotubes when $A/\lambda^2 =1$, $\sigma\eta_0=0.754+5.278\ii$, and the normalized distance between the two tubes is 1.4 ($D/(2\alpha)=1.4$); see Fig.~\ref{fig:Area1PointColormap}. The symmetric parametric model presented in \S\ref{sec:pmodels} with SCP=3,4,5,6 is used. Note here that we assign a separate set of SCPs for each nanotube shape with each SCP being controlled by two parameters, i.e., radius and angle as shown in  \ref{eq:pmodel1}. Therefore, the total number of parameters used in shape optimization is equal to $2\times2\times\mathrm{SCP}$.  The optimized nanotube designs along with the electric field, $|E|^2$, in their interiors are depicted in Fig.~\ref{fig:p1_Results}.

The maximum achieved improvement, i.e., the ratio of the field concentration for the optimized pair over the value corresponding to the circular nanotube pair for the same operating condition, surpasses 32  and it is achieved for 6 SCPs (see Fig.~\ref{fig:p1_SCP6}) with only negligible improvements obtained for $\mathrm{SCPs}>6$ when the symmetric parametric model is used. One may observe that in all cases in Fig.~\ref{fig:p1_Results}, the left nanotube acts as the main ``receiver'' while the right nanotube plays the role of a ``reflector''. As SCPs increase both the receiver and reflector undergo a ``dimerization'' transformation that leads to a layout that resembles a combination of two closely-packed quasi-triangular nanotubes for the receiver and two similarly-shaped nanotubes for the reflector, which are however placed perpendicularly with respect to the first set. Note that the electromagnetic field does not penetrate at all into certain pieces of the scatterers (especially the left one). That feature does not render them useless for the overall signal concentration; indeed, their presence does not increase the collected power internally to the nanotube but influences the energy flow externally to them. In this way, it affects the signal into other spatial regions of the configuration and certainly makes an indispensable part of the sustained resonances.

\begin{figure}[hbtp]
 \centering
  \subfloat[3 SCPs; see Fig.~\ref{fig:p1_SCP3}.]
  {\includegraphics[width = 0.5\textwidth]{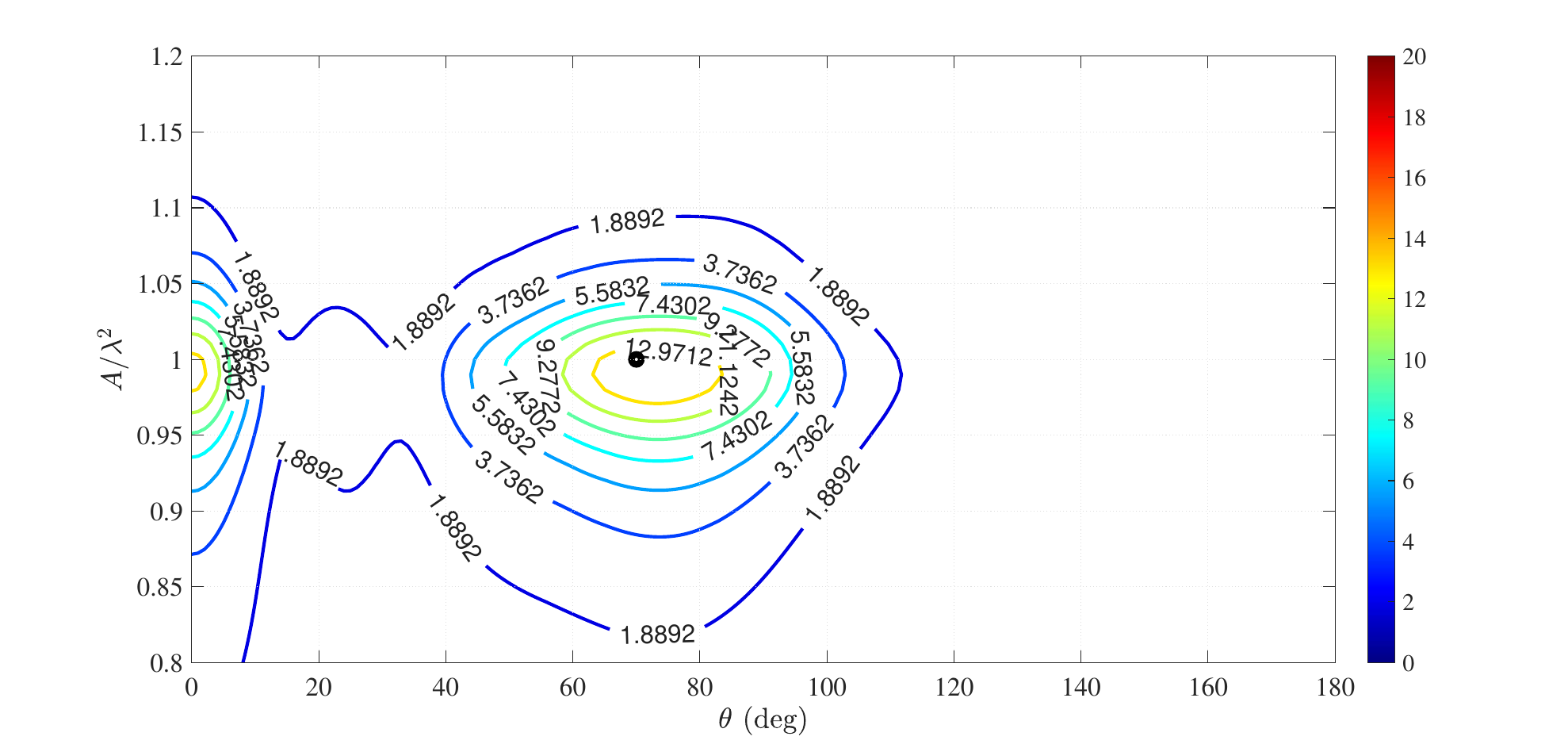}} 
  \subfloat[4 SCPs; see Fig.~\ref{fig:p1_SCP4}.]{\includegraphics[width = 0.5\textwidth]{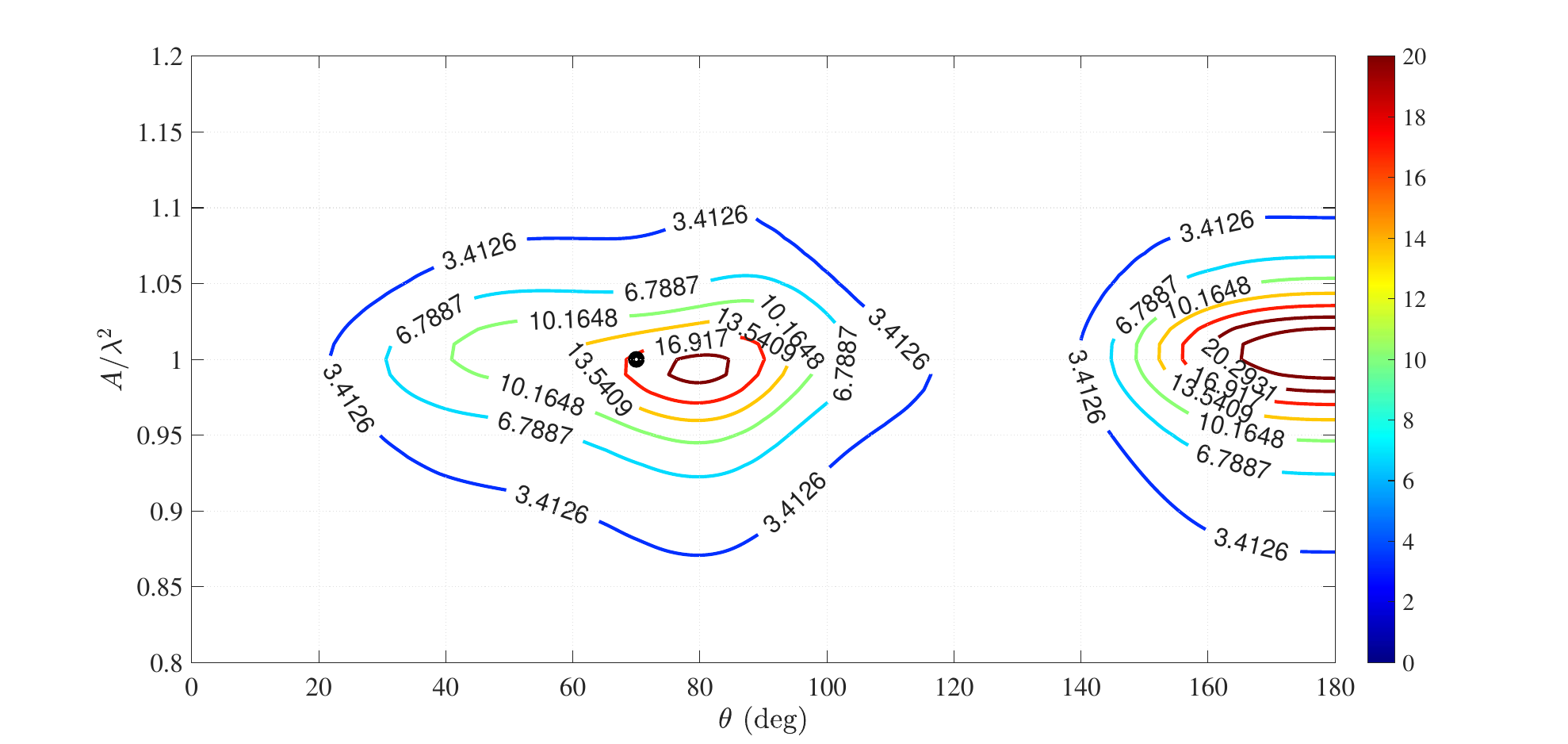}}  \\
  \subfloat[5 SCPs; see Fig.~\ref{fig:p1_SCP5}.]
  {\includegraphics[width = 0.5\textwidth]{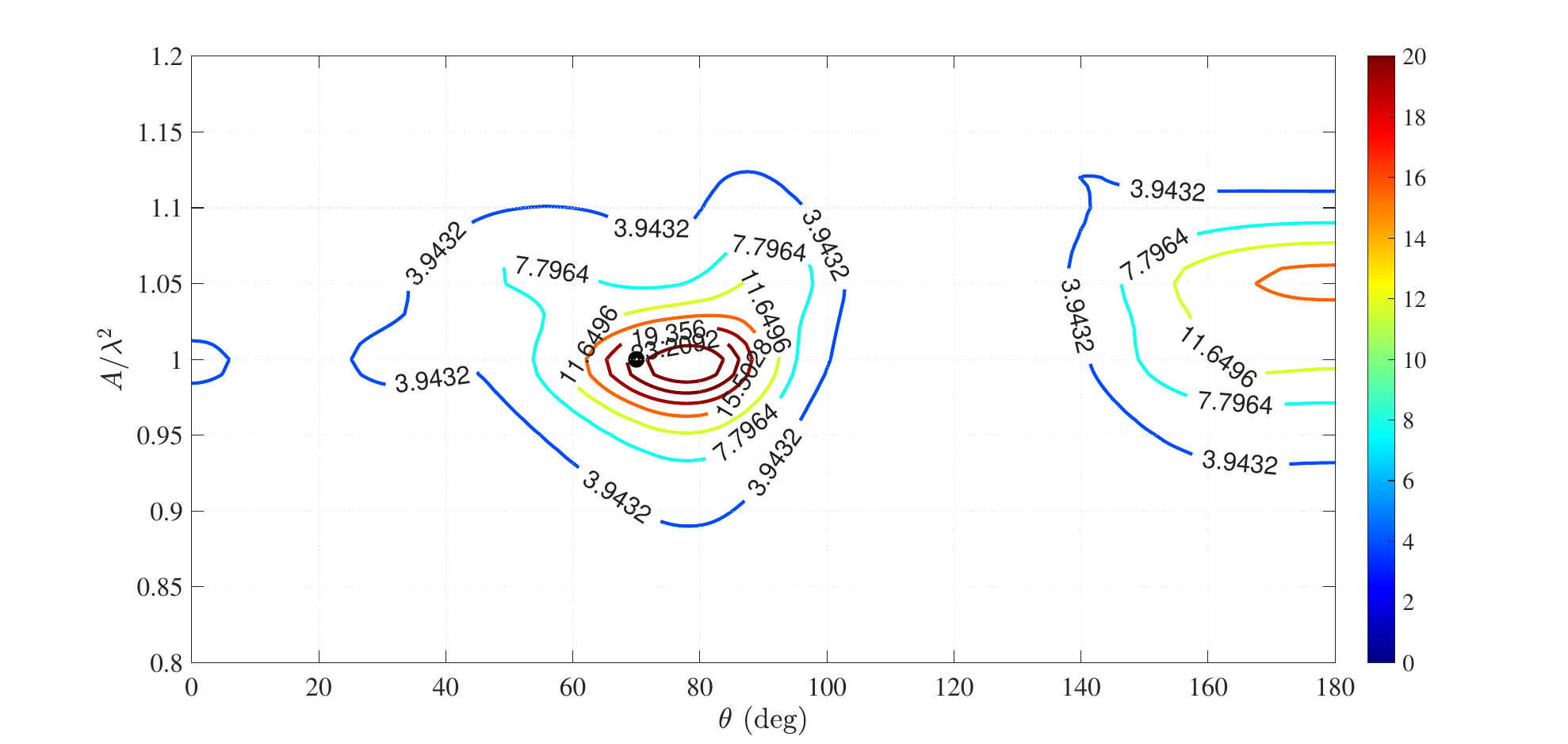}} 
  \subfloat[6 SCPs; see Fig.~\ref{fig:p1_SCP6}.]{\includegraphics[width = 0.5\textwidth]{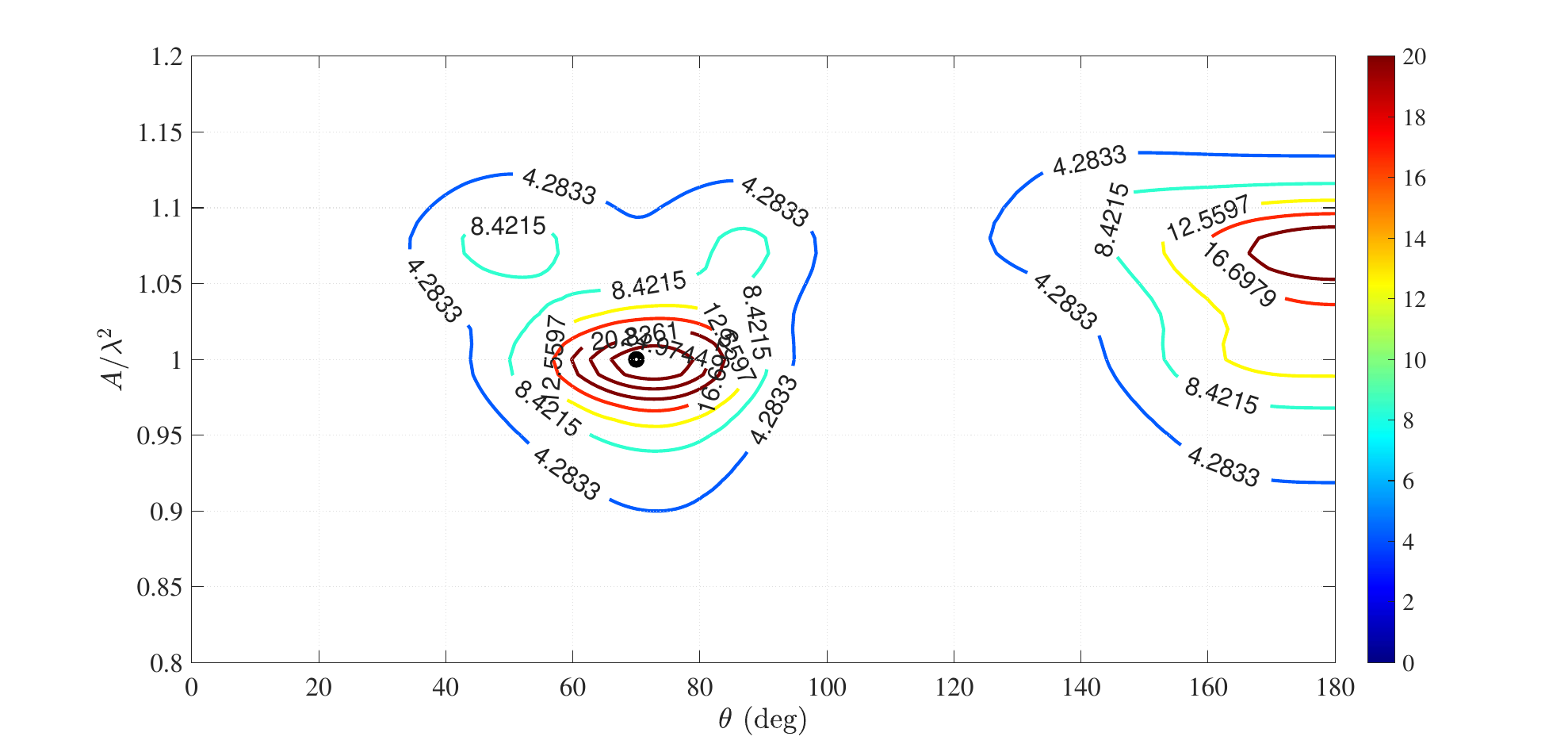}}  
  \caption{Performance robustness of optimized designs of Case I for $A/\lambda^2\in[0.8,1.2]$ and wave angle $\theta\in[0^{\degree},180^{\degree}]$; The nominal operational condition corresponds to the black circle on each of the contour plots above.}
  \label{fig:sensitivity_p1}
\end{figure}

Furthermore, in Fig.~\ref{fig:sensitivity_p1}, the performance of each optimized design shown in Fig.~\ref{fig:p1_Results} is plotted for an extended operational regime within $\pm20\%$ of the nominal $A/\lambda^2$ ratio; the angle $\theta\in[0^\degree,180^\degree]$, as dictated by the new symmetries of the geometries. In all scenarios, the optimized designs exhibit improved performance for a large wave angle range, i.e., $\theta\in[30\degree,100\degree]$, and an $A/\lambda^2$ ratio which is within $\pm10\%$ of the nominal one. This result quantifies the sensitivity of the optimized designs with respect to changes in the EM wavelength and angle, and indicates the attained robustness of these results. 

\begin{figure}[htbp]
    \centering
     \subfloat[3 SCPs]{\includegraphics[height = 0.29\textwidth]{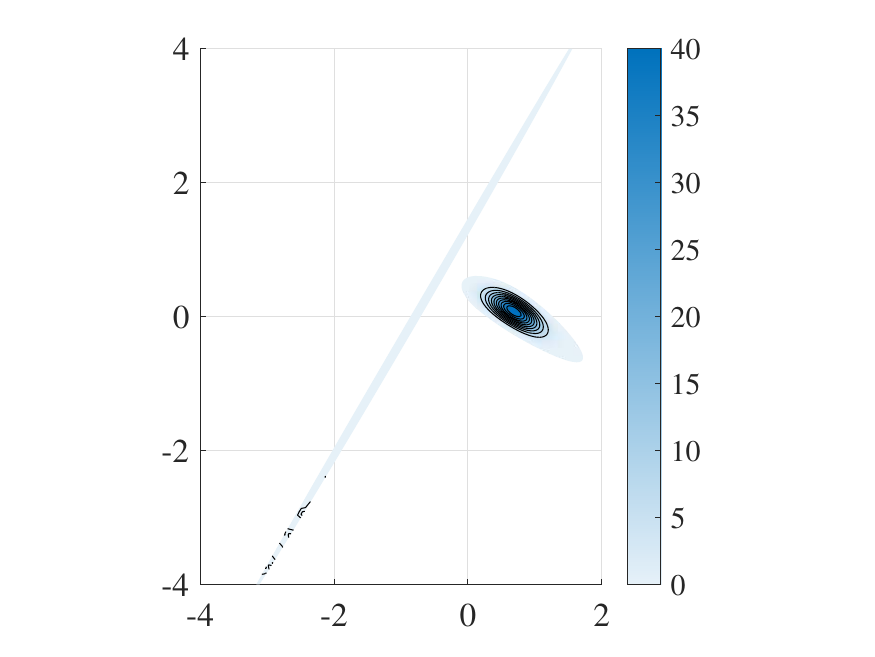}}
    \subfloat[3 SCPs: symmetric vs non-symmetric optimal design\label{fig:p1_sym_nsym_comp}]{\includegraphics[height = 0.29\textwidth]{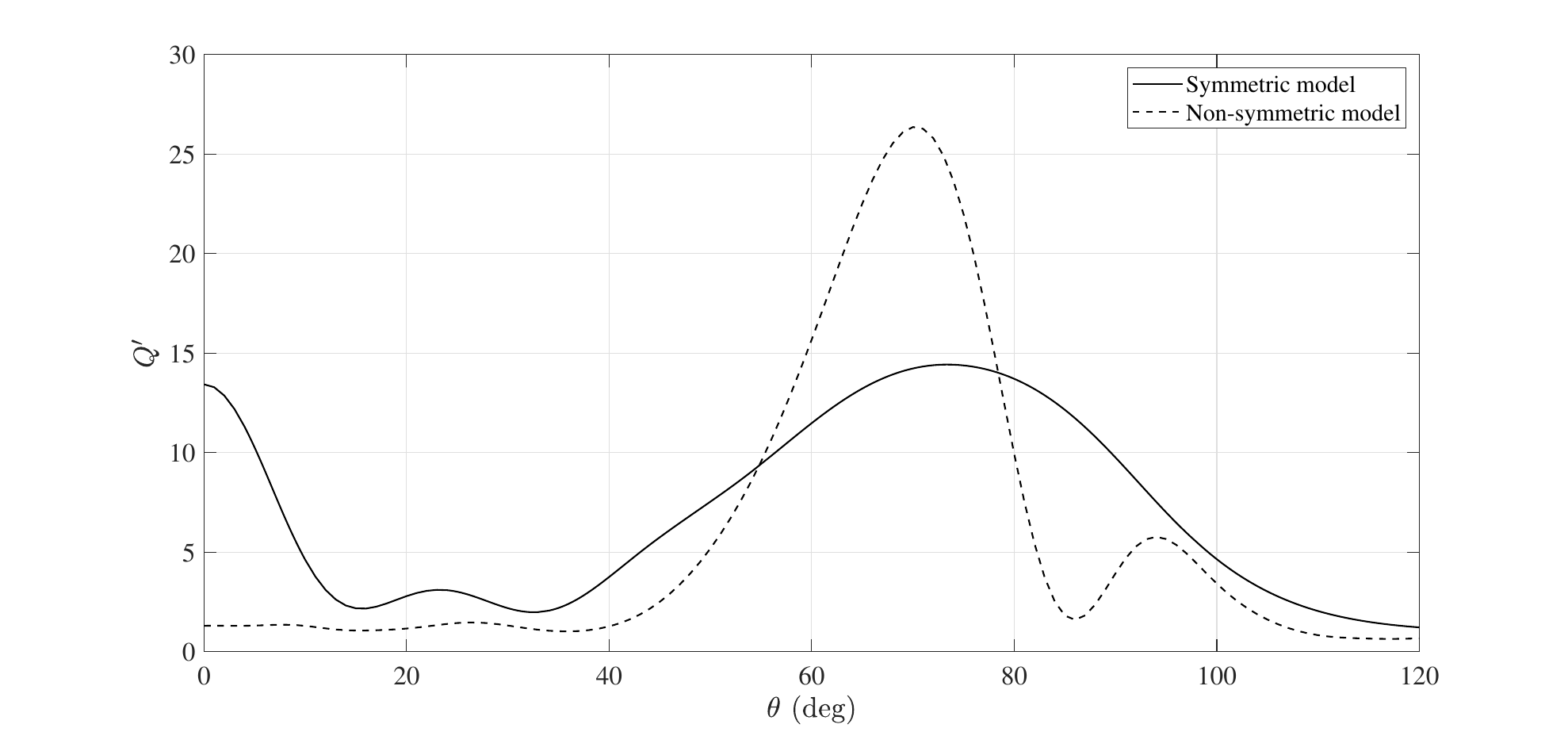}}
    \caption{Optimization results of Case I for 3 SCPs with use of the non-symmetric parametric model. (a) Shape of the design. (b) Relative field enhancement $Q'$ with respect to incoming angle $\theta$ for symmetric and asymmetric model.}
  \label{fig:p1_nsym_Results}
\end{figure}

The same series of optimizations can be performed by employing the non-symmetric parametric model described in \S\ref{sec:pmodels}. Using the non-symmetric parametric model, higher improvements can be generally achieved for the same number of SCPs as anticipated due to the increased number of degrees of freedom and shown in Fig.~\ref{fig:p1_nsym_Results}. Specifically, when using the non-symmetric parametric model, the field concentration improvement is approximately 27 times higher than the circular nanotube pair value, which is significantly larger than what can be achieved with 3 SCPs for the symmetric model. However, the non-symmetric model may occasionally suffer from an ``alignment problem'' which results in orienting elongated nanotube shapes with tiny curvature radii \cite{OnSmoothening} according to the selected wave direction value and consequently reducing the angular robustness of the optimized shape's performance; see  Fig.~\ref{fig:p1_sym_nsym_comp}. This is not necessarily happening for all optimal designs produced by the non-symmetric model, and if one is mainly interested in the highest possible improvement, the non-symmetric model will generally achieve higher values with a potential penalty reducing the design's performance with respect to the wave angle. Finally, if improved performance over a large number of wave angles or wavelengths is an application requirement, the objective function in  \eqref{eq:shape_optmization} can be modified to include a weighted average of multiple wave angles and/or wavelengths.

\begin{figure}[thbp]
 \centering
  \subfloat[3 SCPs, $Q' = 10.39$.\label{fig:p2_SCP3}]{\includegraphics[width = 0.45\textwidth]{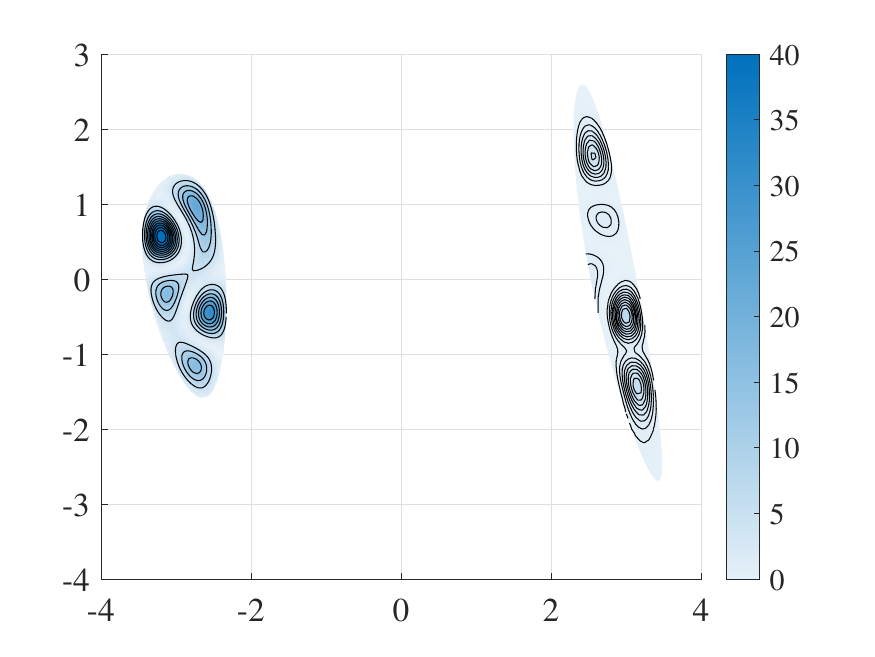}} \hfill
  \subfloat[4 SCPs, $Q' = 14.98$.\label{fig:p2_SCP4}]{\includegraphics[width = 0.45\textwidth]{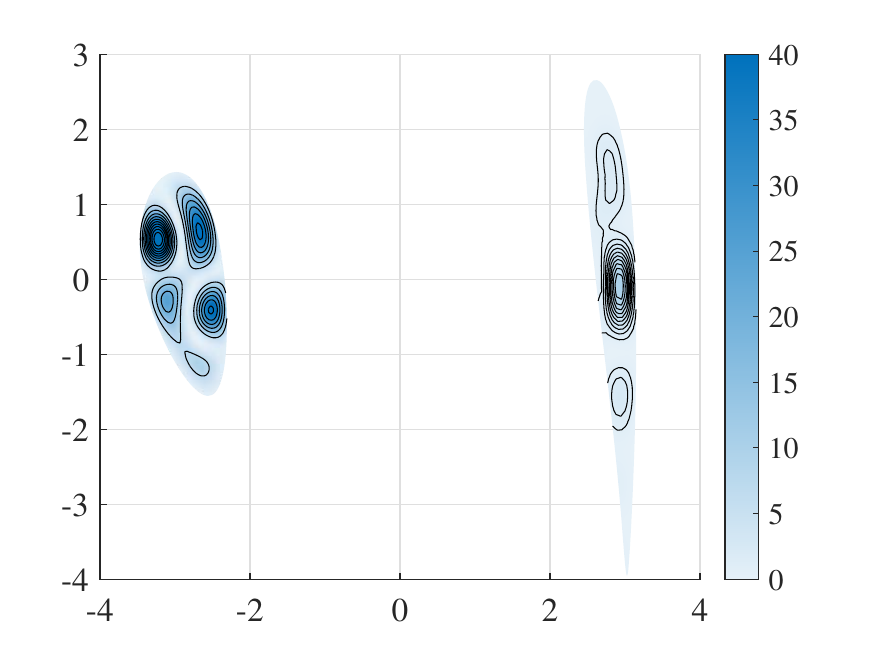}}\\
  \subfloat[5 SCPs, $Q' = 19.88$.\label{fig:p2_SCP5}]{\includegraphics[width = 0.45\textwidth]{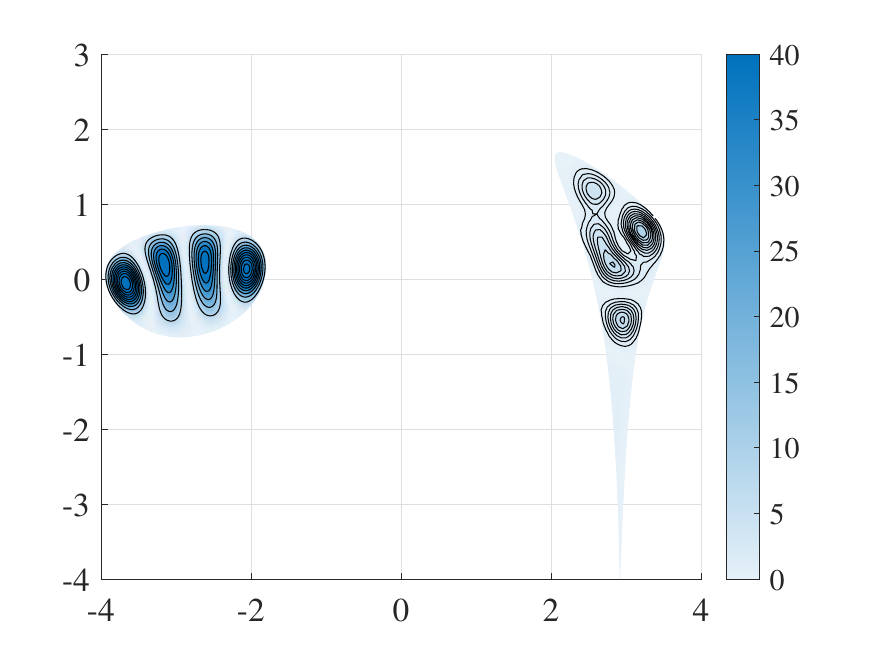}} \hfill
  \subfloat[9 SCPs, $Q' = 23.22$.\label{fig:p2_SCP9}]{\includegraphics[width = 0.45\textwidth]{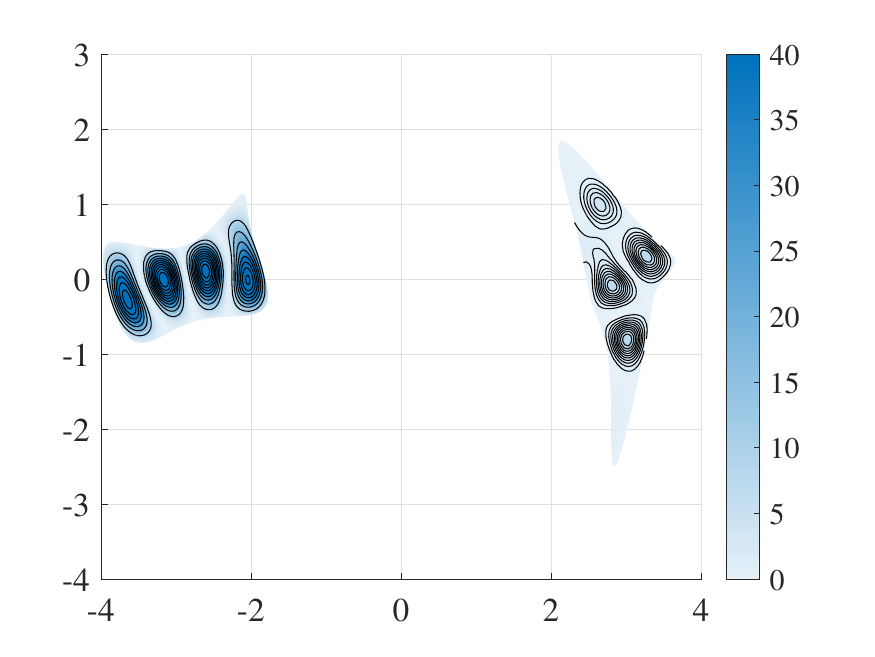}}
  \caption{Optimization results for the Case II with an increasing number of SCPs (3,4,5, and 6) at \mbox{$A/\lambda^2=2.5$}, $D/(2\alpha)=1.4$, $\theta=70{\degree}$, and $\sigma\eta_0=0.754+5.278\ii$, using the ssymmetric parametric model.}
  \label{fig:p2_Results}
\end{figure}

\begin{figure}[htb]
 \centering
  \subfloat[3 SCPs; see Fig.~\ref{fig:p2_SCP3}.]
  {\includegraphics[width = 0.5\textwidth]{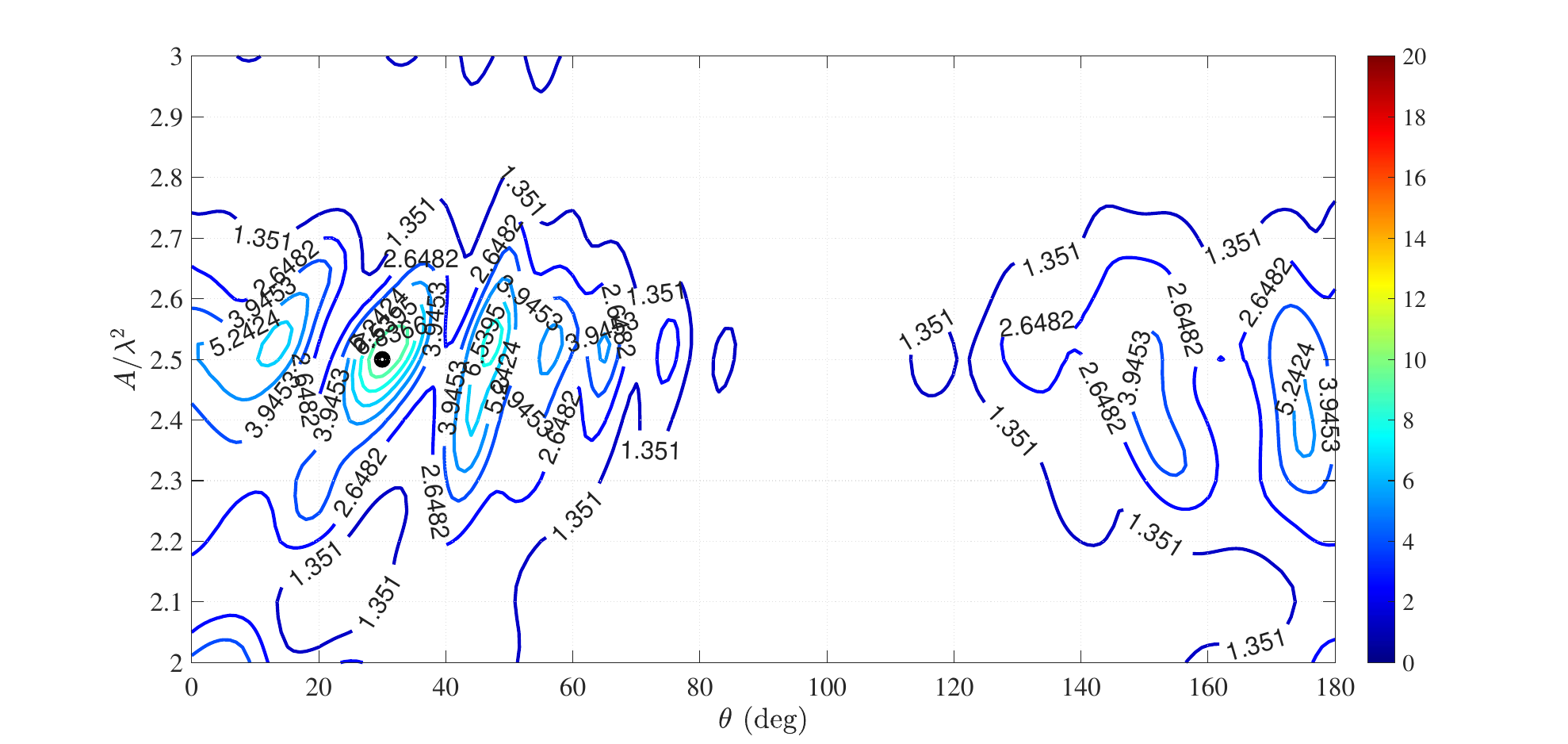}} 
  \subfloat[4 SCPs; see Fig.~\ref{fig:p2_SCP4}.]{\includegraphics[width = 0.5\textwidth]{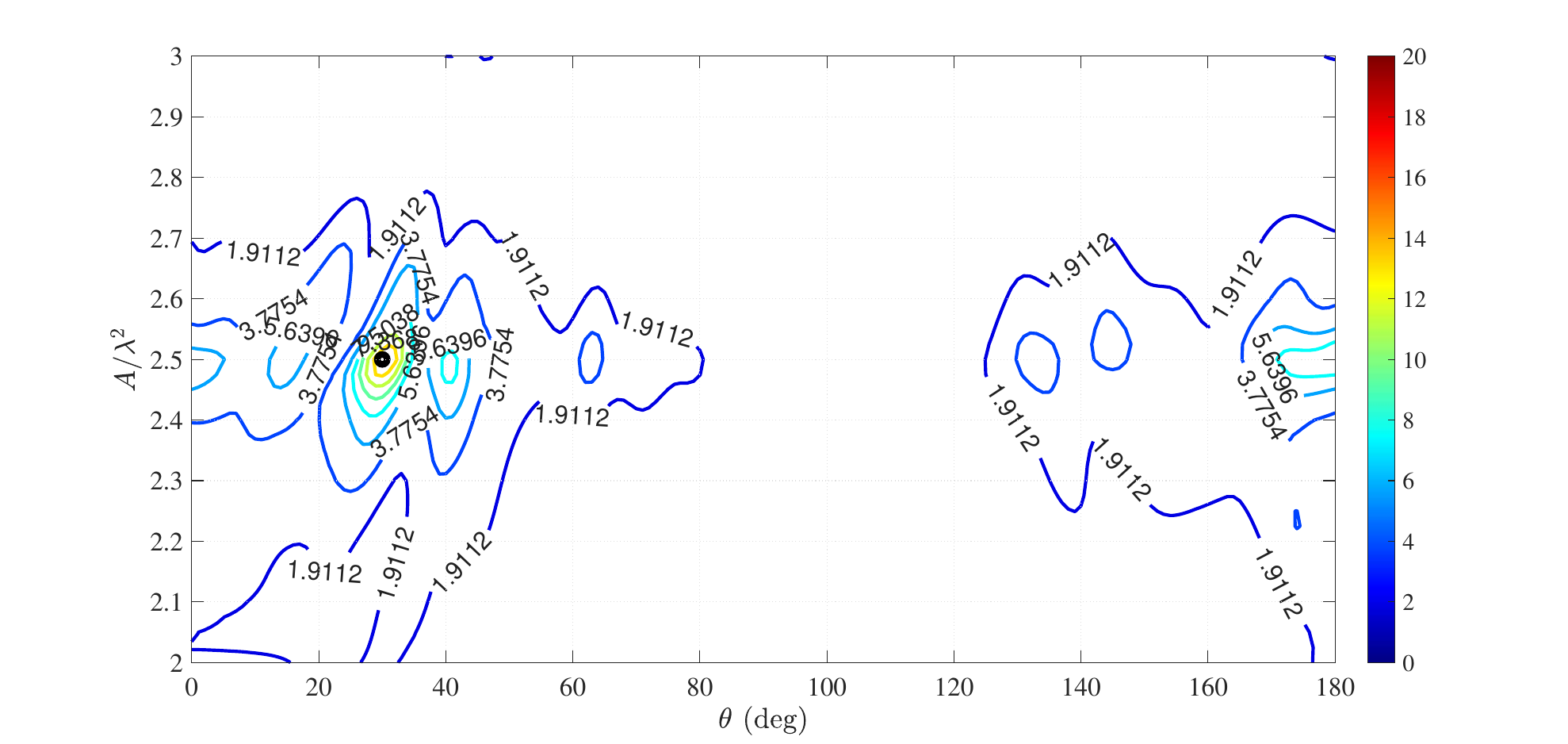}}  \\
  \subfloat[5 SCPs; see Fig.~\ref{fig:p2_SCP5}.]
  {\includegraphics[width = 0.5\textwidth]{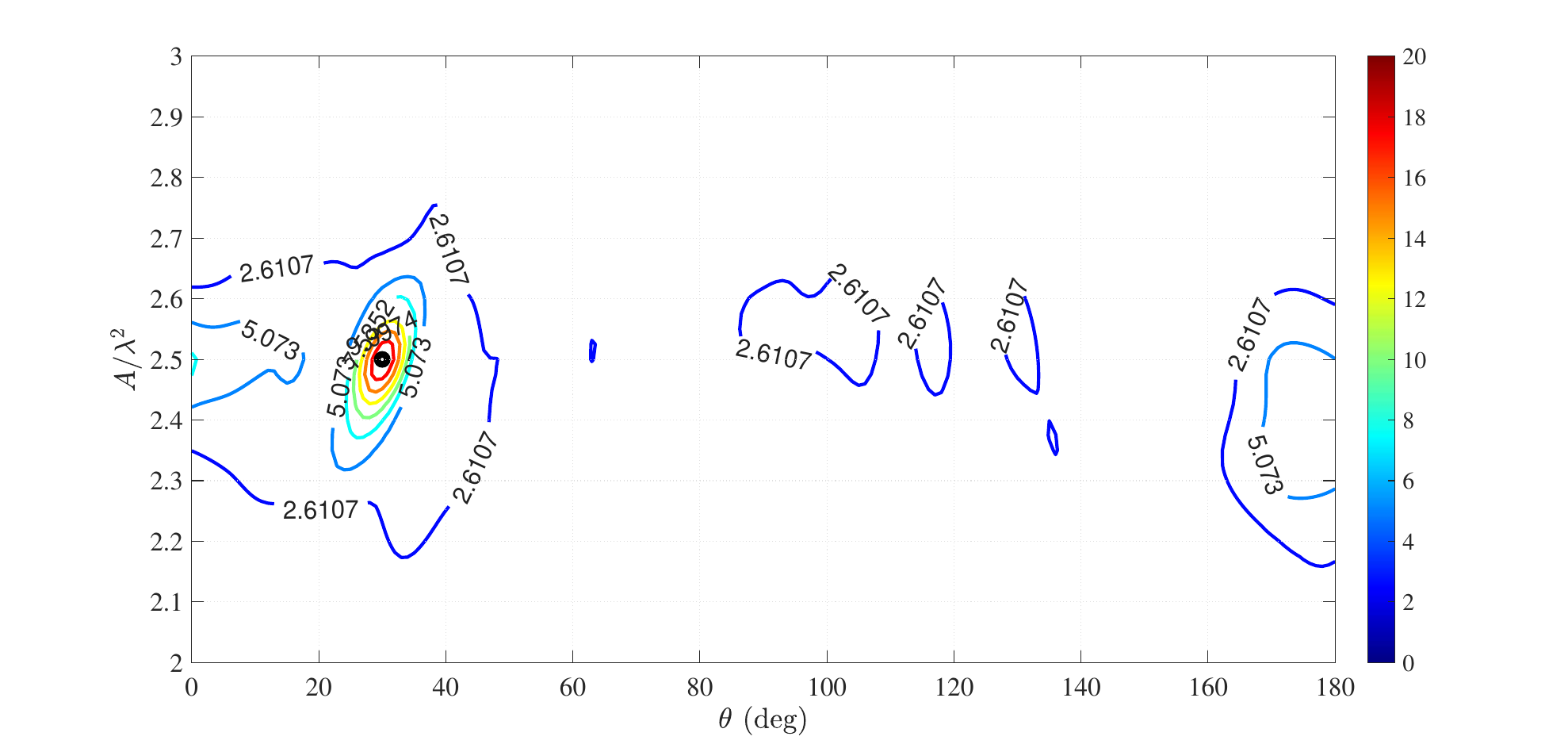}} 
  \subfloat[9 SCPs; see Fig.~\ref{fig:p2_SCP9}.]{\includegraphics[width = 0.5\textwidth]{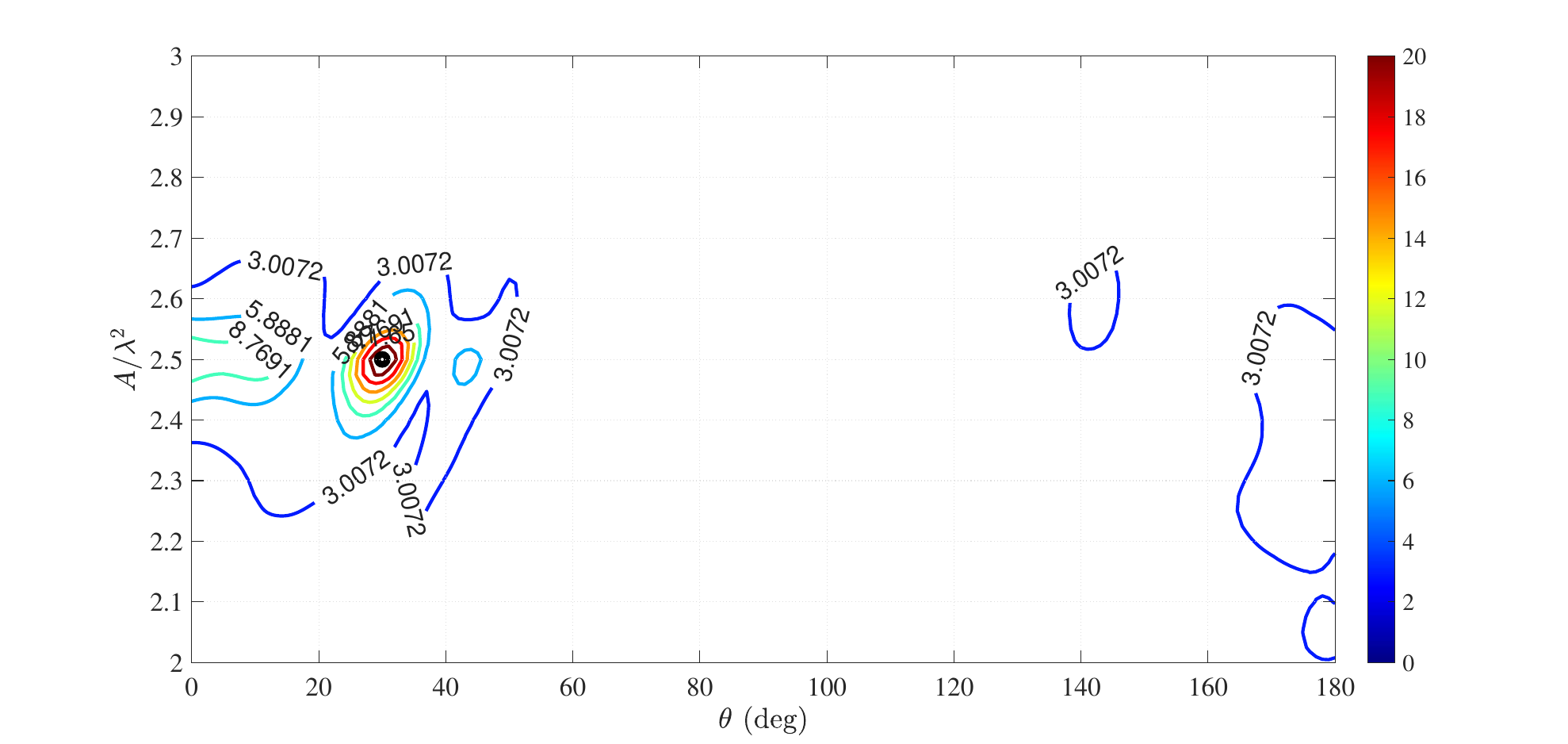}}  
  \caption{Performance robustness of optimized designs of Case II for $A/\lambda^2\in[2,3]$ and wave angle $\theta\in[0^{\degree},180^{\degree}]$; The nominal operational condition corresponds to the black circle on each of the contour plots above.}
  \label{fig:sensitivity_p2}
\end{figure}

For the second example of the Table~\ref{tab:opt conditions}, i.e., Case II, we present the obtained optimal shapes for an increasing number of SCPs in Fig.~\ref{fig:p2_Results} with the response of the optimal designs to variations of the $A/\lambda^2$ ratio and wave angle being depicted in Fig.~\ref{fig:sensitivity_p2}. As already discussed in the presentation of Case I, the number of SCPs is increased until no further improvement (or only negligible improvement) is recorded. The achieved field concentration improvement for the optimal design with 9 SCPs in Case II is 23 times higher than that of the circular nanotube pair; see Fig.~\ref{fig:p2_SCP9}.

The ``dimerization'' behavior exhibited in Case I does not extend to the second example where the $A/\lambda^2$ ratio is increased. Specifically, in Case II, the non-symmetric model achieves significantly better results, although relatively concentrated to a small region around the operational point. However, the improvement, when compared to Case I, is less pronounced and this is mainly due to an increase in the $A/\lambda^2$ ratio  that leads to more peaks for the field within the nanotubes' interior and therefore reduces the impact of the nanotube shape; see Figs.~\ref{fig:p2_Results}. Additionally, the reduced robustness of the non-symmetric model is evident when observing the $A/\lambda^2$ and $\theta$ responses in Fig.~\ref{fig:sensitivity_p2}.

The selected cases have demonstrated that for relatively small $A/\lambda^2$ ratios, the increase reaches values of approximately 30 times better when compared to the circular counterparts while maintaining an improved performance in a relatively large region around the operational point. The achieved improvement is decreased when larger $A/\lambda^2$ ratios are considered with a simultaneous deterioration of the robustness of the performance for slightly varying wave angles or wavelengths. 

\begin{figure}[htbp]
 \centering
  \subfloat[$\widetilde{Q}(D/(2\alpha),\theta)$ for lossless plasmonic nanotubes of moderate optical area.\label{fig:p5_region}]{\includegraphics[width = 0.50\textwidth]{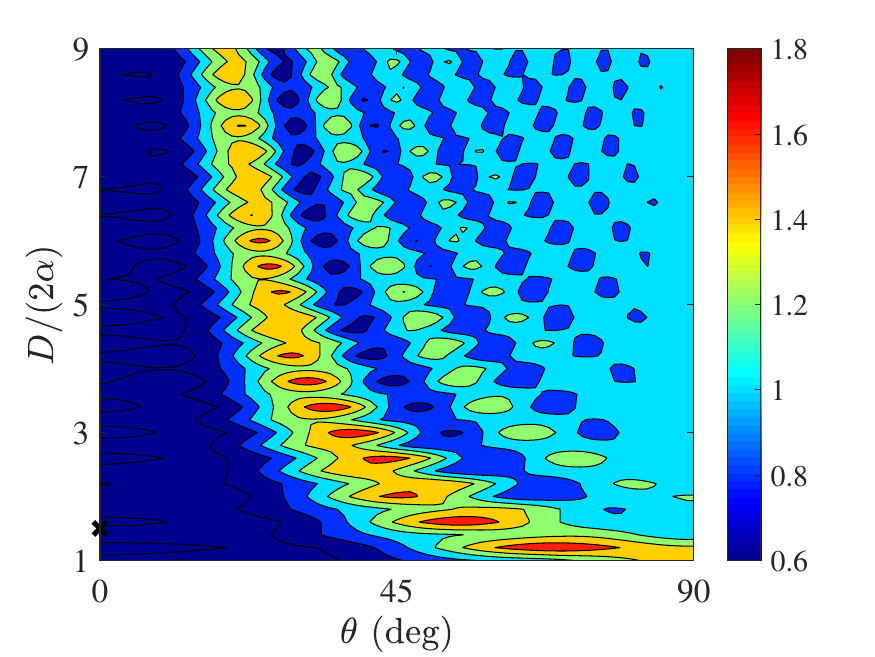}} \\
  \subfloat[5 SCPs, $Q' = 87.56$.\label{fig:p5_SCP5}]{\includegraphics[width = 0.50\textwidth]{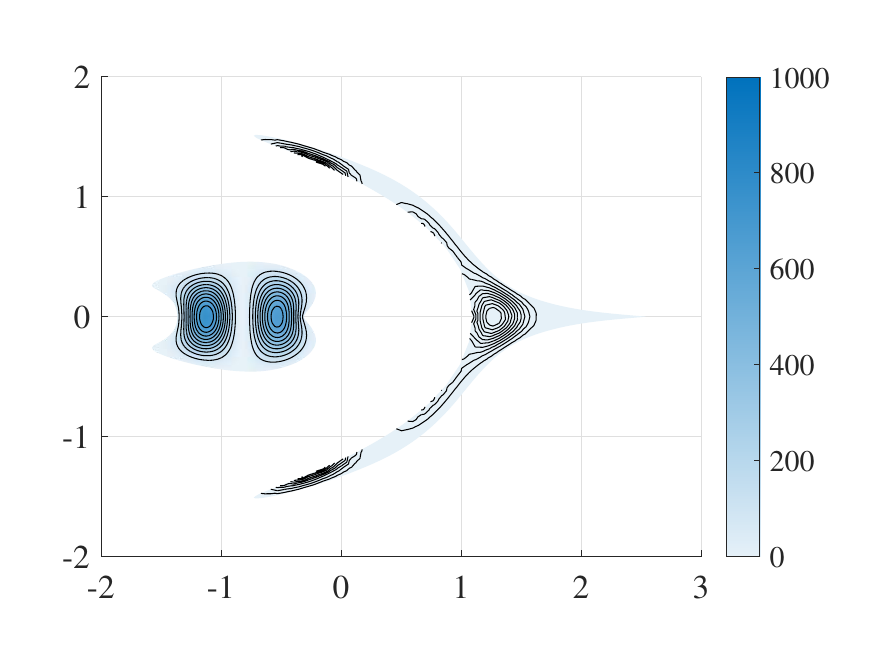}}
  \subfloat[6 SCPs, $Q' = 182.90$.\label{fig:p5_SCP6}]{\includegraphics[width = 0.50\textwidth]{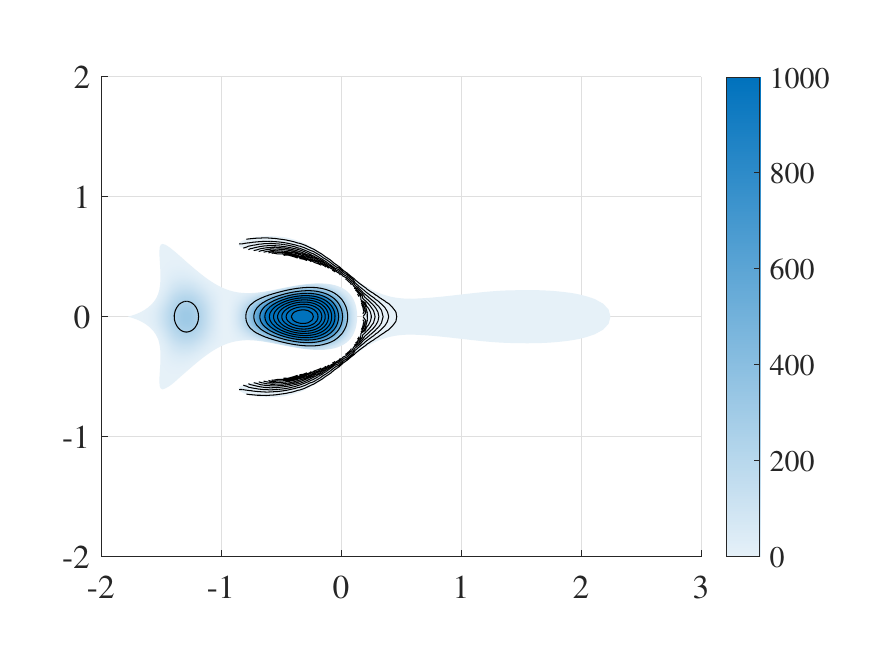}}\\
  \caption{Optimization results for a scenario where the respective circular nanotubes underperform ($A/\lambda^2=1$, $D/(2\alpha)=1.5$, $\theta=0^{\degree}$, and $\sigma\eta_0=0-4.524\ii$). Selected point shown with a black $\bm{\times}$ marker on Fig.~\ref{fig:p5_region}.}
  \label{fig:p5_Results}
\end{figure}

Although the results presented so far were limited in  regions where circular nanotube pairs were good performers, we additionally present here, for reasons of completeness, an indicative shape optimization example in a region where the circular nanotube pair underperforms. Specifically, in Fig.~\ref{fig:p5_region}, we present the operational region, i.e., $A/\lambda^2=1$, $D/(2\alpha)=1.5$, $\theta=0^{\degree}$, and $\sigma\eta_0=0-4.524\ii$ at the black $\bm{\times}$ mark depicted on the same figure. The field concentration improvement goes significantly higher, with an almost 183 relative increase over the circular counterpart when 6 SCPs are used; see Fig.~\ref{fig:p5_SCP6}. Obviously, this result can only be generalized for operational regimes where circular nanotube pairs deliver poorly and, therefore, put the performance bar at particularly low level .

\section{Conclusion}\label{sec:conclusion}
The present work develops a methodology for the computation of the electric field across nanotube pairs illuminated by electromagnetic waves, using an IsoGeometric-Analysis-based Boundary Element Method (IGABEM); it demonstrably achieves high accuracy with a relatively low number of degrees of freedom. We employ this approach in studying the electric field concentration within circular nanotube pairs in close proximity and comparing these values with two circular nanotubes at an infinite distance. A systematic investigation over a wide range of area over structural, textural and source characteristics has been conducted and presented. Subsequently, regions in which circular nanotubes already exhibit high field concentrations are identified and selected for nanotube-shape optimization. Specifically, for such scenarios, we manage to produce arbitrary-shaped nanotubes which achieve field concentrations exceeding by at least 30 times the corresponding circular nanotube pair's concentration. This performance improvement becomes significantly higher, 180 times or more, when cases of underperforming circular nanotube pairs are selected. The optimized designs maintain their performance over a large region of wave angles and variations of the cross section areas of nanotubes. 

The proposed method can be easily extended to treat a finite number of nanotubes in arrays. Being able to reliably solve metasurfaces of a finite number of nanotubes with various distances between them and each one with differently shaped cross section, may offer unprecedented capabilities in the inverse design and optimization of similar setups. Although infinite gratings with identical meta-atoms are analytically solvable \cite{FloquetBloch} with use of the Floquet-Bloch theorem, the efficient treatment of finite electromagnetic metasurfaces constitutes an open research challenge with large scientific and industrial interest. The proposed methodology, along with the insights gained by the performed shape optimizations, e.g., ``dimerization'' trends, can pave the way for developing an appropriate toolbox for the efficient design of metasurfaces with significant applications in green devices \cite{Tsiftsis2022} and analog signal processing \cite{FieldTransformation}.

\section*{Acknowledgments}
This work has received funding from Nazarbayev University, Kazakhstan under the Collaborative Research Grant Project: ``Waves Interaction with CArbon Nanotubes - WICAN'', Grant Award Nr. 20122022CRP1607.

\bibliographystyle{elsarticle-num}
\bibliography{main}

\newpage\appendix
\label{AppendixA}
\section{Pairs of Identical Circular Nanotubes: Performance Limits for the Simplest Possible Coupled Design}
\label{app1}
\setcounter{figure}{0}
\begin{figure}[!thbp]
   \centering
   
   \subfloat[$\underset{\theta}{\max}\left(\widetilde{Q}(\Im(\sigma),D/2\alpha)\right)$ for $A/\lambda^2= 0.5$\label{fig:p_study_app0.5}]{\includegraphics[width = 0.48\textwidth]{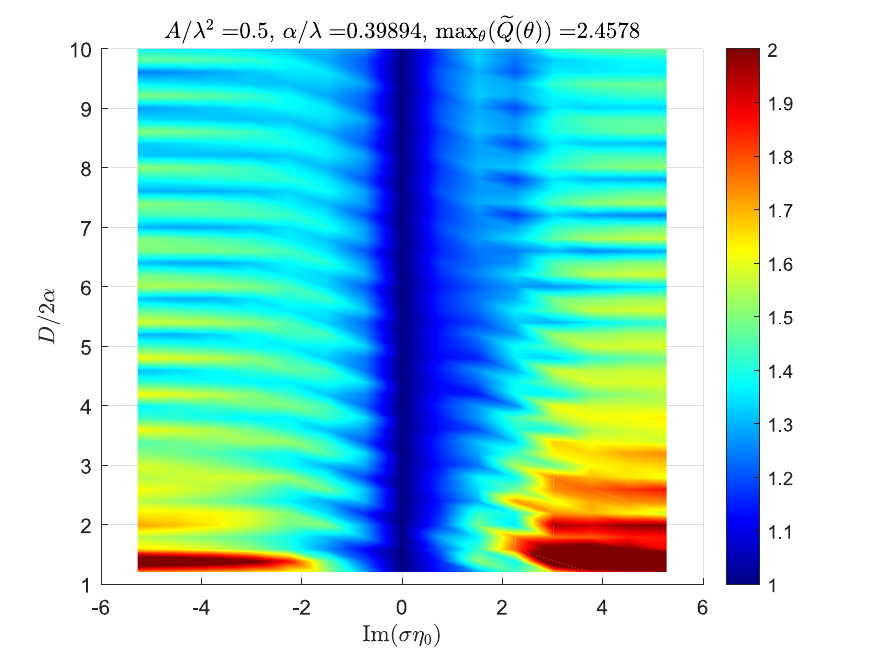}} \quad
   \subfloat[$\underset{\theta}{\max}\left(\widetilde{Q}(\Im(\sigma),D/2\alpha)\right)$ for $A/\lambda^2= 1$\label{fig:p_study_app1}]{\includegraphics[width = 0.48\textwidth]{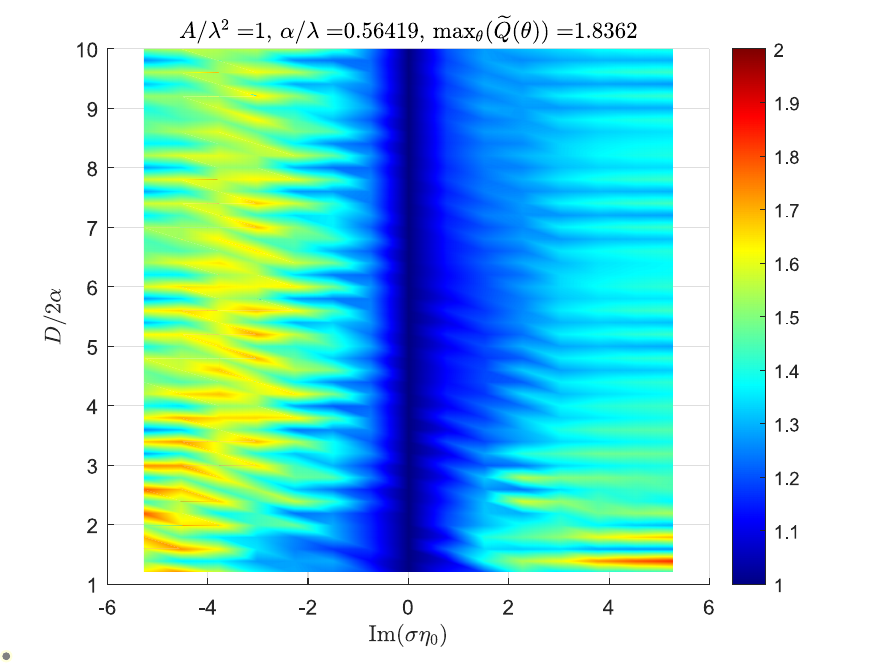}}
   
   \subfloat[$\underset{\theta}{\max}\left(\widetilde{Q}(\Im(\sigma),D/2\alpha)\right)$ for $A/\lambda^2= 1.5$\label{fig:p_study_app1.5}]{\includegraphics[width = 0.48\textwidth]{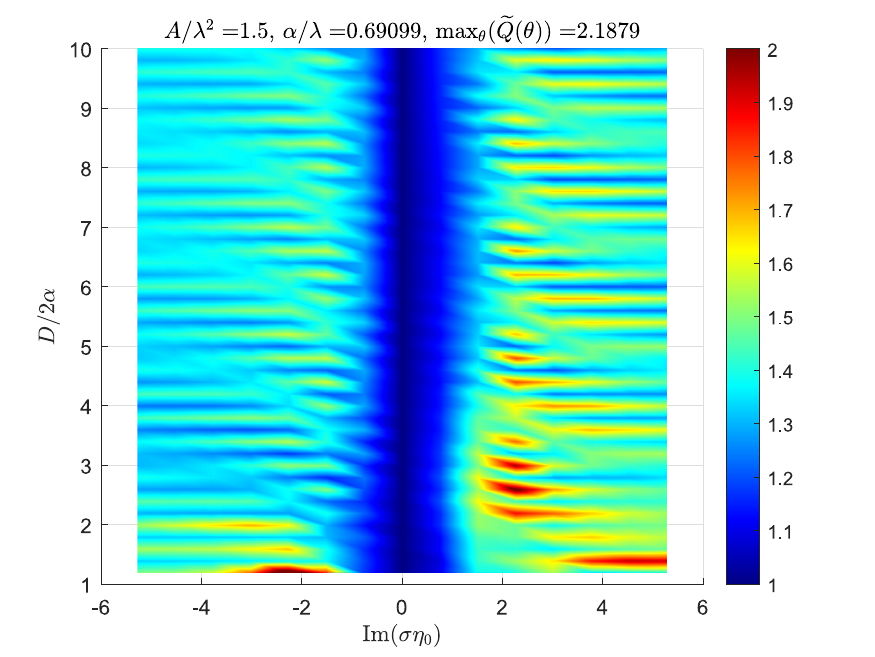}}\quad
   \subfloat[$\underset{\theta}{\max}\left(\widetilde{Q}(\Im(\sigma),D/2\alpha)\right)$ for $A/\lambda^2= 2$\label{fig:p_study_app2}]{\includegraphics[width = 0.48\textwidth]{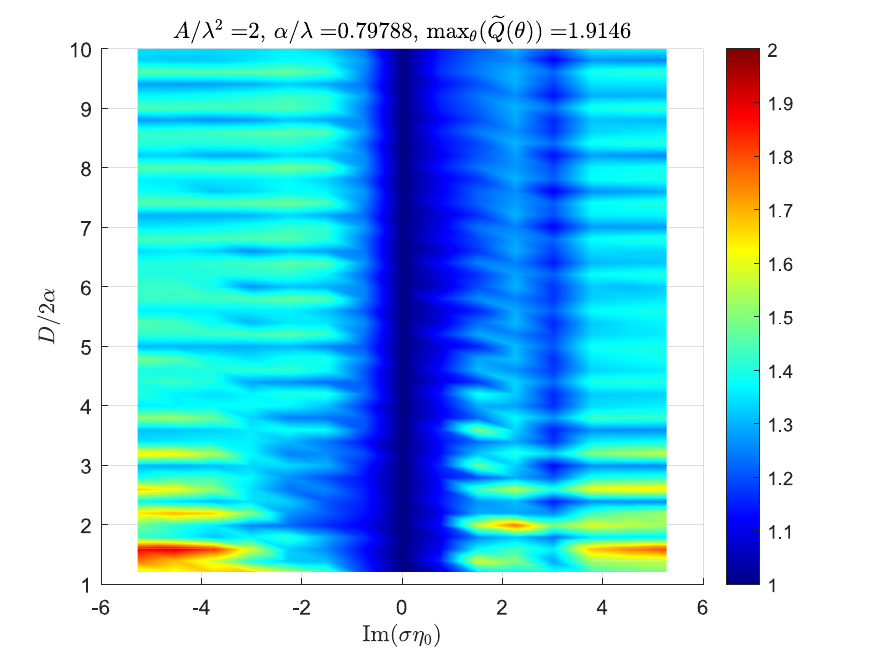}}

   \caption{Maxima of the normalized electric field concentration, $\underset{\theta}{\max}(\widetilde{Q})$, with respect to $\Im(\sigma\eta_0)$ and $D/2\alpha$ for $A/\lambda^2 \in [0.5,2]$.}
  \label{fig:p_study_1}
\end{figure}

\begin{figure}[thbp]
   \centering
    \subfloat[$\underset{\theta}{\max}\left(\widetilde{Q}(\Im(\sigma),D/2\alpha)\right)$ for $A/\lambda^2= 2.5$\label{fig:p_study_app2.5}]{\includegraphics[width = 0.48\textwidth]{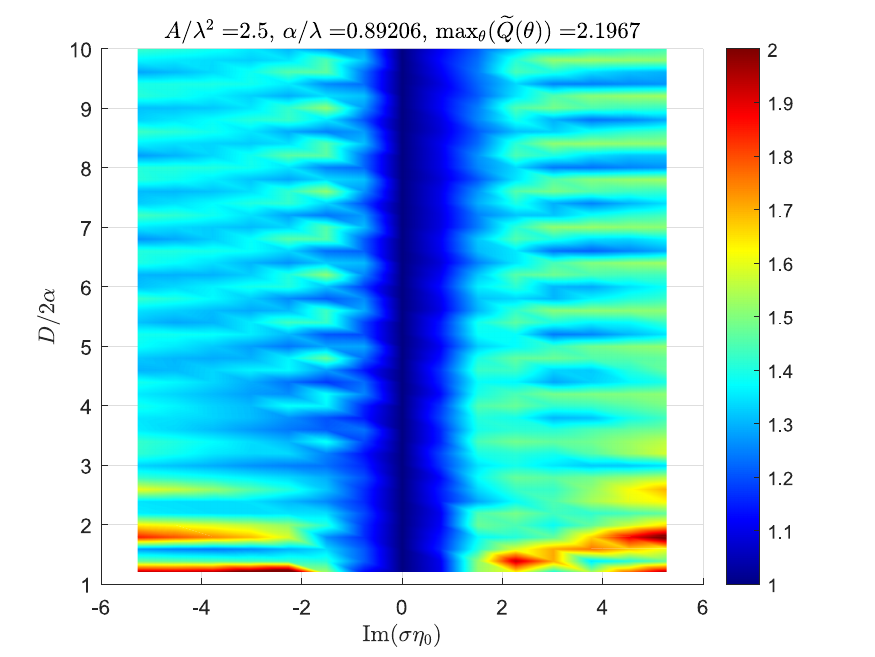}}\quad
   \subfloat[$\underset{\theta}{\max}\left(\widetilde{Q}(\Im(\sigma),D/2\alpha)\right)$ for $A/\lambda^2= 3$\label{fig:p_study_app3}]{\includegraphics[width = 0.48\textwidth]{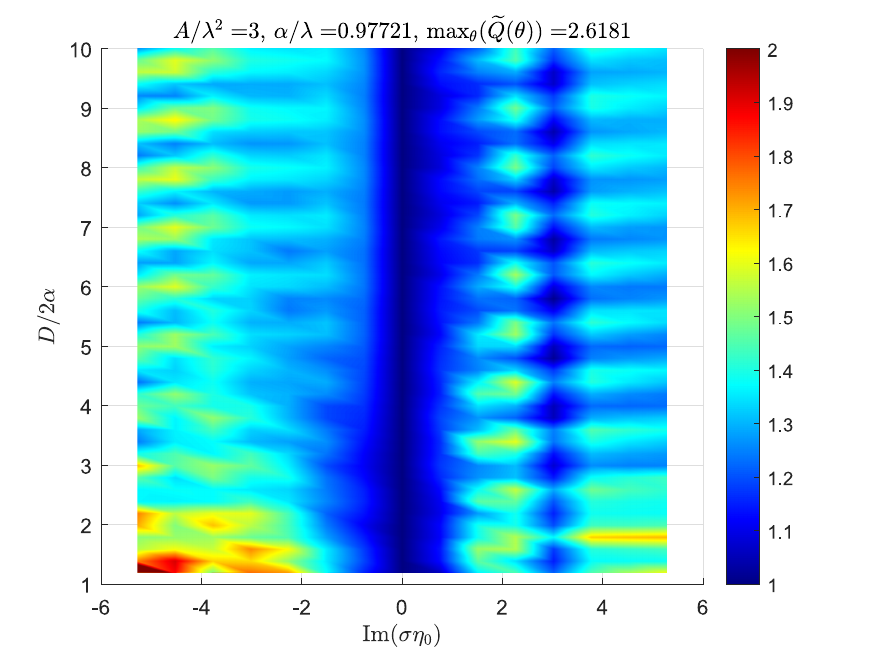}}
   
   \subfloat[$\underset{\theta}{\max}\left(\widetilde{Q}(\Im(\sigma),D/2\alpha)\right)$ for $A/\lambda^2= 3.5$\label{fig:p_study_app3.5}]{\includegraphics[width = 0.48\textwidth]{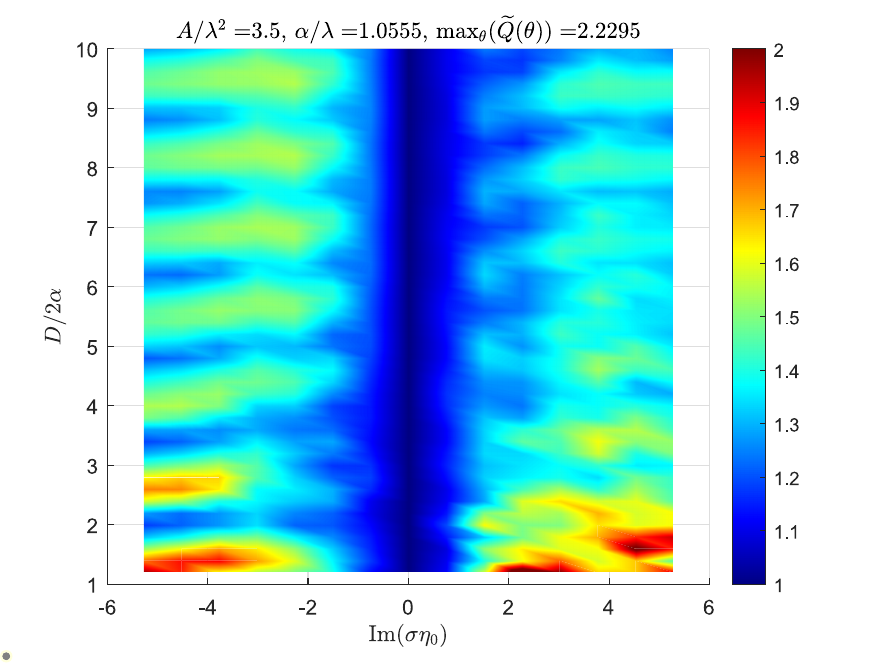}}\quad
   \subfloat[$\underset{\theta}{\max}\left(\widetilde{Q}(\Im(\sigma),D/2\alpha)\right)$ for $A/\lambda^2= 4$\label{fig:p_study_app4}]{\includegraphics[width = 0.48\textwidth]{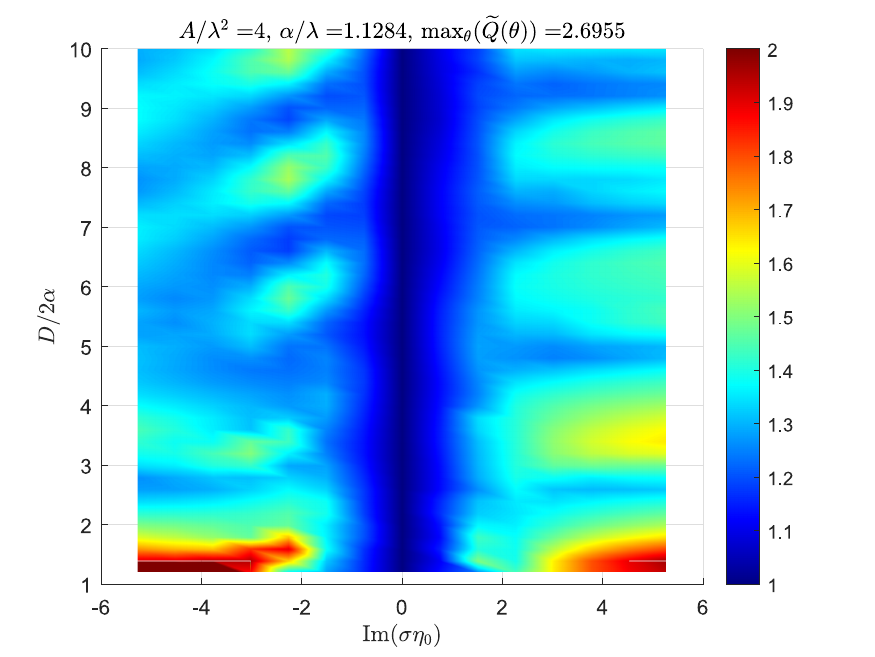}}
   
   \subfloat[$\underset{\theta}{\max}\left(\widetilde{Q}(\Im(\sigma),D/2\alpha)\right)$ for $A/\lambda^2= 4.5$\label{fig:p_study_app4.5}]{\includegraphics[width = 0.48\textwidth]{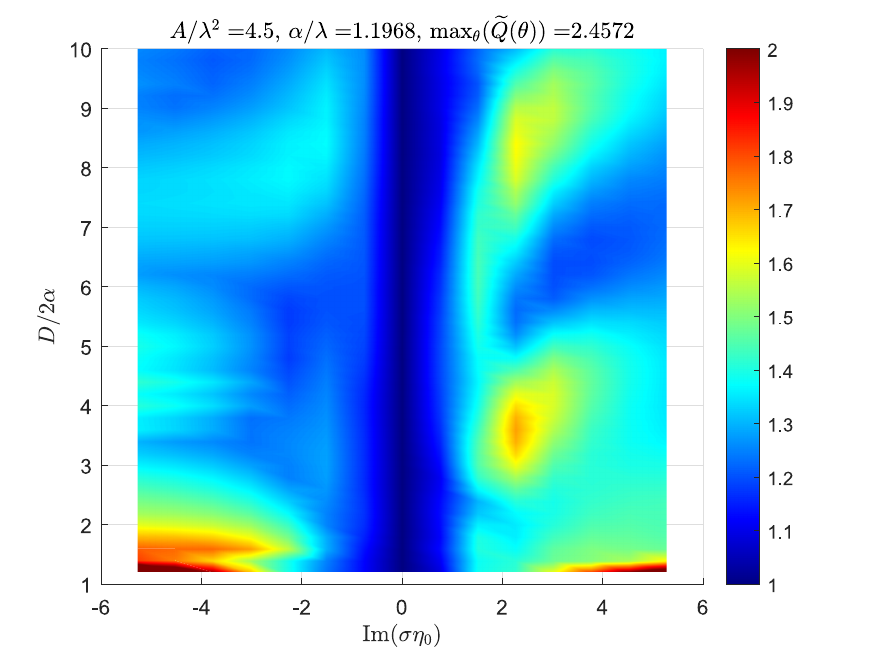}} \quad
   \subfloat[$\underset{\theta}{\max}\left(\widetilde{Q}(\Im(\sigma),D/2\alpha)\right)$ for $A/\lambda^2= 5$\label{fig:p_study_app5}]{\includegraphics[width = 0.48\textwidth]{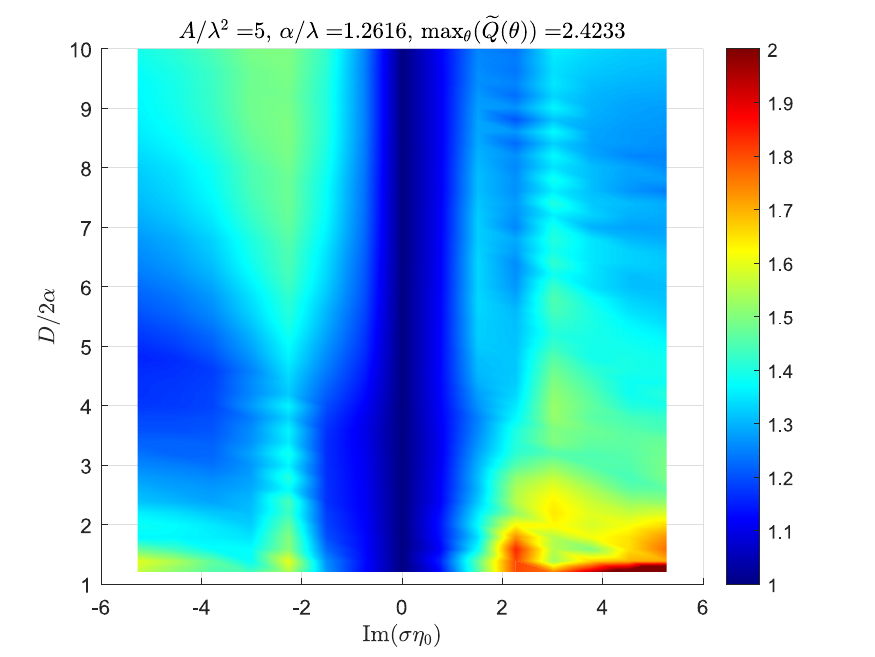}}

   \caption{Maxima of the normalized electric field concentration, $\underset{\theta}{\max}(\widetilde{Q})$, with respect to $\Im(\sigma\eta_0)$ and $D/2\alpha$ for $A/\lambda^2 \in [2.5,5]$.}
  \label{fig:p_study_2}
\end{figure}

\begin{figure}[thbp]
   \centering

\subfloat[$\underset{\theta}{\max}\left(\widetilde{Q}(\Im(\sigma),D/2\alpha)\right)$ for $A/\lambda^2= 5.5$\label{fig:p_study_app5.5}]{\includegraphics[width = 0.48\textwidth]{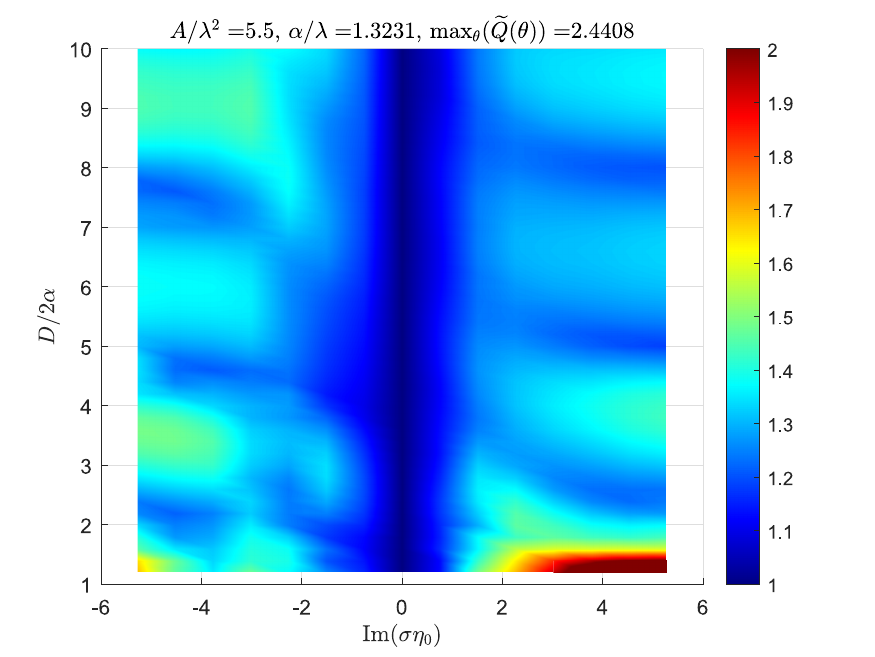}} \quad
   \subfloat[$\underset{\theta}{\max}\left(\widetilde{Q}(\Im(\sigma),D/2\alpha)\right)$ for $A/\lambda^2= 6$\label{fig:p_study_app6}]{\includegraphics[width = 0.48\textwidth]{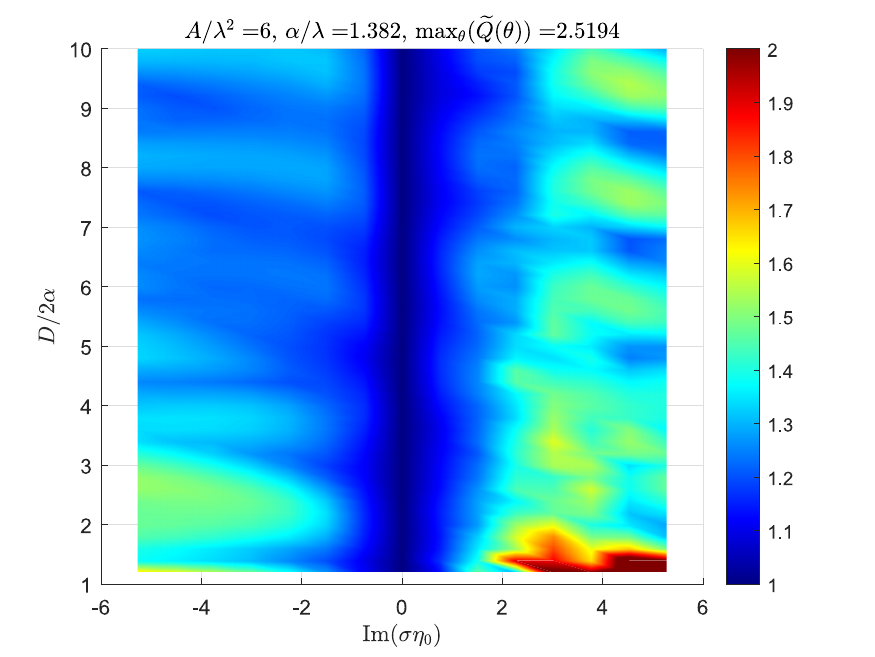}}
   
    \subfloat[$\underset{\theta}{\max}\left(\widetilde{Q}(\Im(\sigma),D/2\alpha)\right)$ for $A/\lambda^2= 6.5$\label{fig:p_study_app6.5}]{\includegraphics[width = 0.48\textwidth]{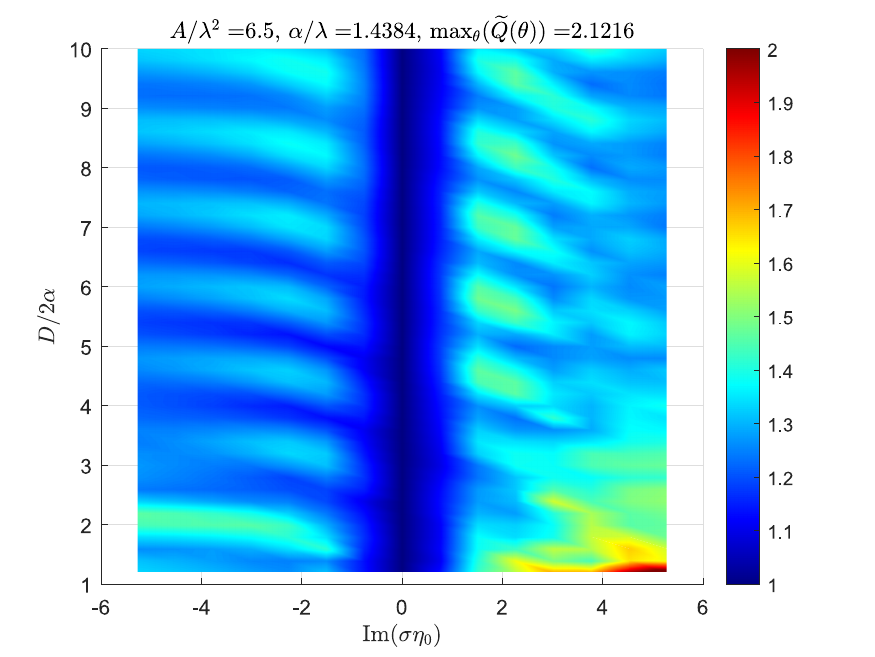}} \quad
   \subfloat[$\underset{\theta}{\max}\left(\widetilde{Q}(\Im(\sigma),D/2\alpha)\right)$ for $A/\lambda^2= 7$\label{fig:p_study_app7}]{\includegraphics[width = 0.48\textwidth]{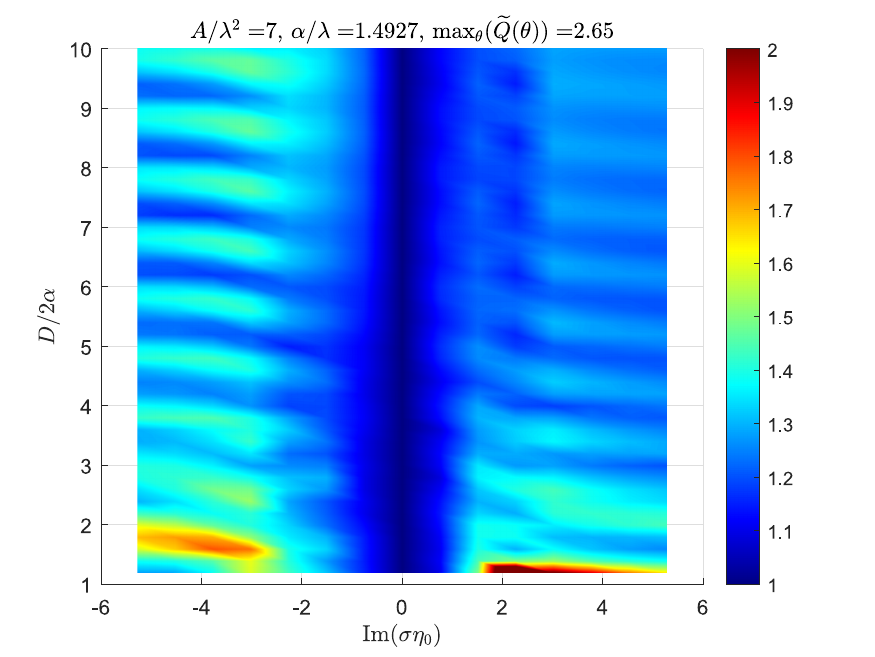}}

    \subfloat[$\underset{\theta}{\max}\left(\widetilde{Q}(\Im(\sigma),D/2\alpha)\right)$ for $A/\lambda^2= 7.5$\label{fig:p_study_app7.5}]{\includegraphics[width = 0.48\textwidth]{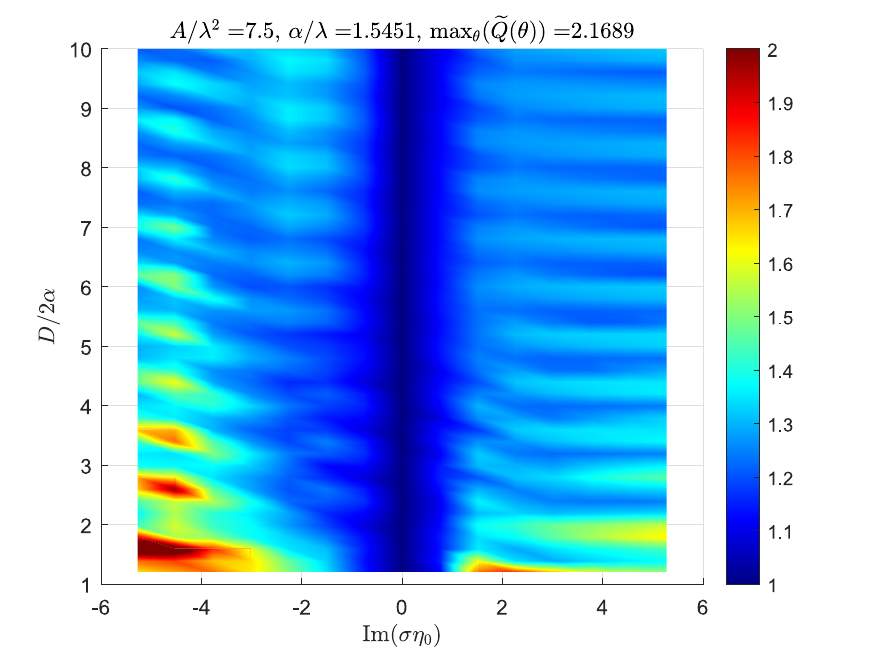}} \quad
   \subfloat[$\underset{\theta}{\max}\left(\widetilde{Q}(\Im(\sigma),D/2\alpha)\right)$ for $A/\lambda^2= 8$\label{fig:p_study_app8}]{\includegraphics[width = 0.48\textwidth]{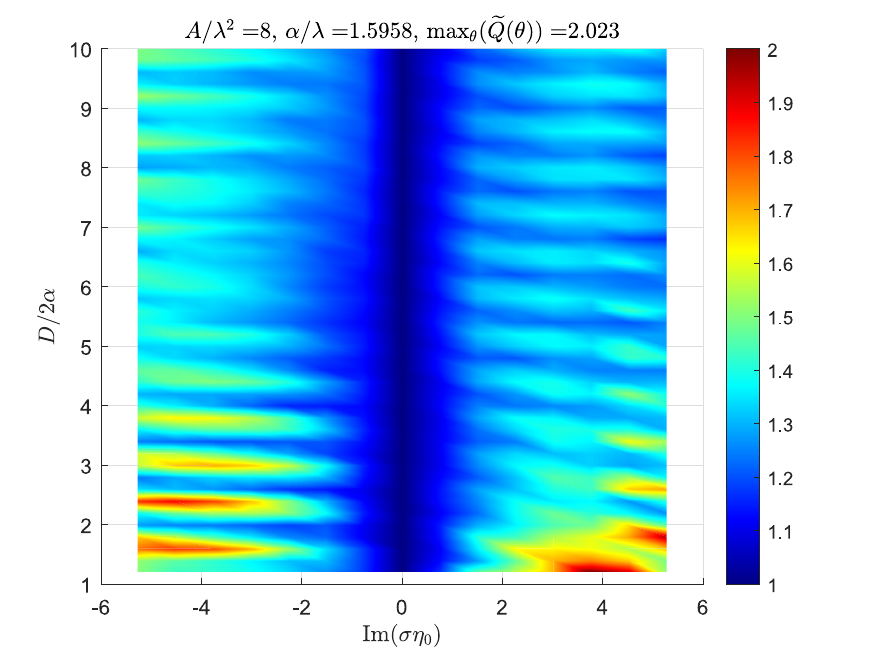}}

   \caption{Maxima of the normalized electric field concentration, $\underset{\theta}{\max}(\widetilde{Q})$, with respect to $\Im(\sigma\eta_0)$ and $D/2\alpha$ for $A/\lambda^2 \in [5.5,8]$.}
  \label{fig:p_study_3}
\end{figure}

\begin{figure}[thbp]
   \centering
  
\subfloat[$\underset{\theta}{\max}\left(\widetilde{Q}(\Im(\sigma),D/2\alpha)\right)$ for $A/\lambda^2= 8.5$\label{fig:p_study_app8.5}]{\includegraphics[width = 0.48\textwidth]{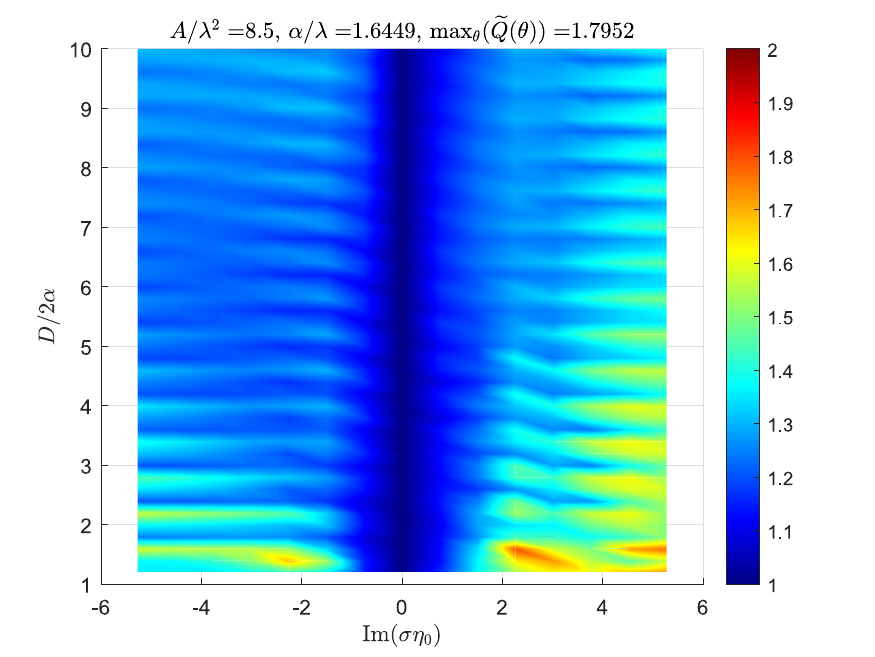}} \quad
   \subfloat[$\underset{\theta}{\max}\left(\widetilde{Q}(\Im(\sigma),D/2\alpha)\right)$ for $A/\lambda^2= 9$\label{fig:p_study_app9}]{\includegraphics[width = 0.48\textwidth]{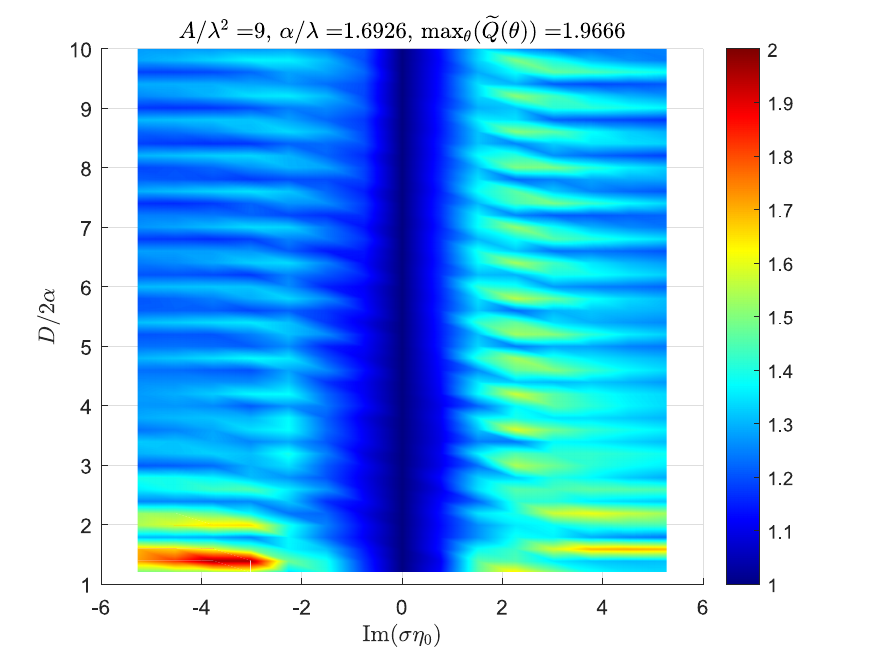}}
   
    \subfloat[$\underset{\theta}{\max}\left(\widetilde{Q}(\Im(\sigma),D/2\alpha)\right)$ for $A/\lambda^2= 9.5$\label{fig:p_study_app9.5}]{\includegraphics[width = 0.48\textwidth]{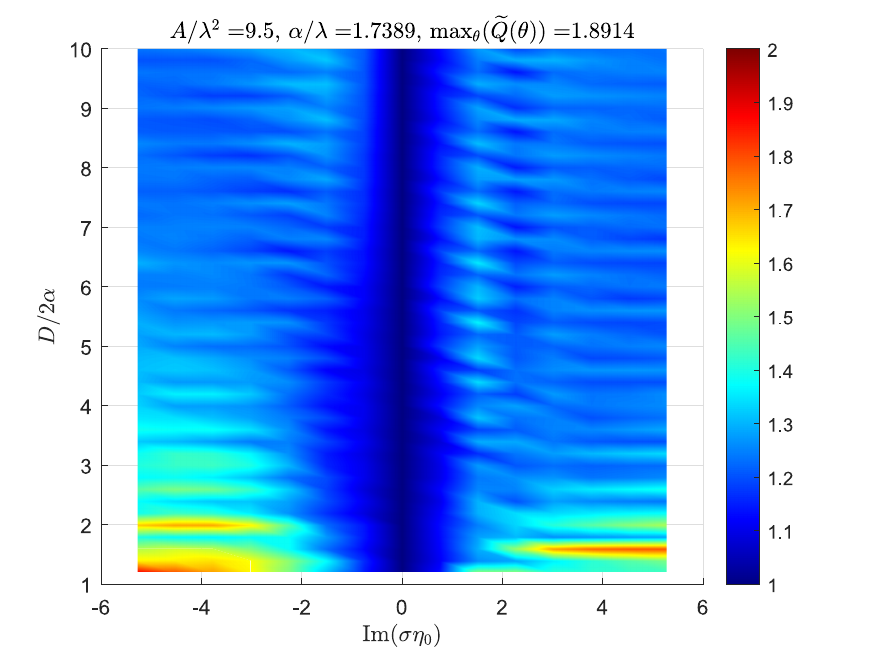}} \quad
   \subfloat[$\underset{\theta}{\max}\left(\widetilde{Q}(\Im(\sigma),D/2\alpha)\right)$ for $A/\lambda^2= 10$\label{fig:p_study_app10}]{\includegraphics[width = 0.48\textwidth]{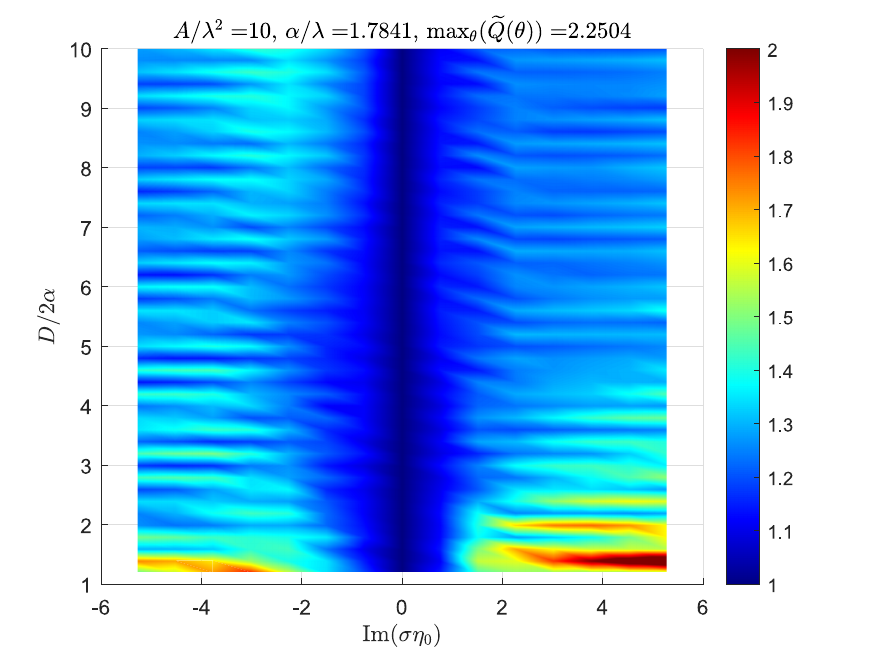}}

   \caption{Maxima of the normalized electric field concentration, $\underset{\theta}{\max}(\widetilde{Q})$, with respect to $\Im(\sigma\eta_0)$ and $D/2\alpha$ for $A/\lambda^2 \in [8.5,10]$.}
  \label{fig:p_study_4}
\end{figure}

%\end{linenumbers}
\end{document}